\documentclass[a4paper,11pt]{article}
\pdfoutput=1
\usepackage{jcappub}
\usepackage[utf8]{inputenc}
\usepackage{amsfonts}
\usepackage{amsmath}
\usepackage{amssymb}
\usepackage{amsthm}
\usepackage{color}
\usepackage{float}
\usepackage[dvipsnames]{xcolor}
\usepackage[normalem]{ulem}
\usepackage{cancel}
\usepackage{graphicx}
\usepackage{appendix}
\usepackage{tabularx, booktabs, multirow}

\newcommand{\be}{\begin{equation}}
\newcommand{\ee}{\end{equation}}

\newcommand{\comment}[1]{} 

\usepackage{scalerel,tikz}
\usetikzlibrary{svg.path}
\definecolor{orcidlogocol}{HTML}{A6CE39}
\tikzset{orcidlogo/.pic={
 \fill[orcidlogocol] svg{M256,128c0,70.7-57.3,128-128,128C57.3,256,0,198.7,0,128C0,57.3,57.3,0,128,0C198.7,0,256,57.3,256,128z};
 \fill[white] svg{M86.3,186.2H70.9V79.1h15.4v48.4V186.2z}
 svg{M108.9,79.1h41.6c39.6,0,57,28.3,57,53.6c0,27.5-21.5,53.6-56.8,53.6h-41.8V79.1z M124.3,172.4h24.5c34.9,0,42.9-26.5,42.9-39.7c0-21.5-13.7-39.7-43.7-39.7h-23.7V172.4z}
 svg{M88.7,56.8c0,5.5-4.5,10.1-10.1,10.1c-5.6,0-10.1-4.6-10.1-10.1c0-5.6,4.5-10.1,10.1-10.1C84.2,46.7,88.7,51.3,88.7,56.8z};
}}
\newcommand\orcidicon[1]{\href{https://orcid.org/#1}{\mbox{\scalerel*{
\begin{tikzpicture}[yscale=-1,transform shape]
\pic{orcidlogo};
\end{tikzpicture}
}{|}}}}

\title{Cosmological Signatures of Curvature-Coupled Dark Energy}

\author[a]{Abdolali Banihashemi~\orcidicon{0000-0003-1107-1253},}
\author[a]{Farbod Hassani~\orcidicon{0000-0003-2640-4460},}
\author[b,c]{Alessandro Casalino~\orcidicon{0000-0001-6709-5292},}
\author[a]{David F. Mota~\orcidicon{0000-0003-3141-142X},}
\author[d,e]{Emilio Bellini~\orcidicon{0000-0003-4762-0795}}
\affiliation[a]{Institute of Theoretical Astrophysics, Universitetet i Oslo, 0315 Oslo, Norway}
\affiliation[b]{
Max Planck Computing and Data Facility, Gießenbachstraße 2, 85748 Garching, Germany
}
\affiliation[c]{
Alma Mater Studiorium - Università di Bologna, via Piero Gobetti 93/2, 40129 Bologna, Italy
}
\affiliation[d]{{Center for Astrophysics and Cosmology}, University of Nova Gorica, Nova Gorica, Slovenia}
\affiliation[e]{IFPU, Institute for Fundamental Physics of the Universe, via Beirut 2, I-34151 Trieste, Italy}

\emailAdd{abdolali.banihashemi@astro.uio.no}
\emailAdd{farbod.hassani@astro.uio.no}
\emailAdd{alessandro.casalino@mpcdf.mpg.de}
\emailAdd{d.f.mota@astro.uio.no}
\emailAdd{emilio.bellini@ung.si}
\abstract{ 
We study a curvature-coupled dark energy model that can modify cosmological
evolution both before recombination and during the late-time accelerated era.
The model belongs to the class of scalar--tensor theories, in which a
quintessence field is non-minimally coupled to the Ricci scalar. We specify
the model through a shifted quartic coupling,
$f(\varphi)=\alpha(\varphi^2-\varphi_{\rm today}^2)^2$,
and an inverse power-law potential,
$V(\varphi)=\Lambda\varphi^{-\sigma}$, where
$\varphi_{\rm today}$ is a constant fixed by requiring the effective Planck mass to
recover its present-day normalization. We implement the model in a
modified version of \texttt{hi\_class} and compute its background and linear
cosmological predictions.

This specific form of non-minimal coupling allows an effective crossing of
the phantom divide at late times while naturally suppressing deviations from standard gravity today and
satisfying local gravity constraints. At the same time, the scalar field
can modify the expansion history before recombination, shifting the acoustic
scale in the direction required to alleviate the $H_0$ tension, while
remaining dynamically relevant as dark energy at low redshift.

For the parameter choices studied here, we find tens-of-percent deviations from
$\Lambda$CDM in the expansion history, matter clustering, and metric-potential
spectra. The modified evolution of the gravitational potentials leaves
characteristic signatures in relativistic observables, with weak-lensing power suppressed by $\mathcal{O}(20$--$40\%)$ at low multipoles and
order-unity changes in the late Integrated Sachs--Wolfe signal. Together,
these results reveal a broad, scale-dependent phenomenology that motivates both a full parameter-space analysis and an extension of the
present linear treatment to a dedicated non-linear $N$-body implementation.
}
\graphicspath{Images/}
\begin{document}
\maketitle

\section{Introduction}

The accelerated expansion of the Universe \cite{SupernovaSearchTeam:1998fmf, SupernovaCosmologyProject:1998vns} remains one of the central open questions in cosmology. The standard cosmological model, $\Lambda$CDM, provides the simplest phenomenological description of this acceleration and is broadly consistent with a wide range of cosmological observations \cite{Planck:2015xua, Planck:2018vyg}. In $\Lambda$CDM, cosmic acceleration is caused by a cosmological constant,
$\Lambda$. Although phenomenologically successful, this term raises two
well-known theoretical problems. The observed value of $\Lambda$ is many
orders of magnitude smaller than naive quantum field theory expectations, a
discrepancy known as the old cosmological constant problem. A second issue,
known as the new cosmological constant problem, concerns the near coincidence
$\rho_\Lambda\sim\rho_m$ at the present epoch. Since
$\rho_m\propto a^{-3}$ while $\rho_\Lambda$ is constant, their ratio evolves
rapidly, making the near equality today occur during a special epoch of the
cosmic expansion history \cite{Weinberg:2000yb}.

From an observational perspective, $\Lambda$CDM has been remarkably successful in reproducing isolated observations.  However, several tensions between different observations have persisted, and in some cases sharpened, as the precision of cosmological data has improved. In particular, the value of $H_0$ inferred from Cosmic Microwave Background (CMB) observations differs from that obtained using late Universe distance measurements, while the amplitude of matter clustering predicted within $\Lambda$CDM is somewhat larger than that inferred from several low redshift probes. In addition, a range of further anomalies has been discussed in the recent literature; see, for example, \cite{CosmoVerseNetwork:2025alb}.
More recently, baryon acoustic oscillation measurements from DESI, especially when combined with CMB and supernova data, have been interpreted as providing hints for a time dependent dark energy sector, with reconstructions that may favour an effective crossing of the phantom divide, $w_{\rm DE}=-1$ \cite{DESI:2025zgx}. While the statistical and theoretical interpretation of this result remains under active discussion, it further motivates models in which the effective dark energy equation of state can depart from that of a cosmological constant.
 Taken together, these indications motivate the exploration of extensions beyond $\Lambda$CDM. At the same time, such models require theoretical predictions of comparable robustness in order to allow meaningful comparisons with observations.

Among the many alternatives to the standard cosmological model, a widely studied direction is to introduce a scalar degree of freedom alongside cold dark matter. Depending on how this scalar couples to gravity, such models can be interpreted either as dark energy extensions or as part of the broader modified gravity landscape. In this work we focus on curvature-coupled dark energy models, corresponding to the class of scalar-tensor theories, formulated within the Horndeski framework \cite{Horndeski:1974wa}.

Standard minimally coupled quintessence relies on a slowly rolling scalar field whose energy density remains nearly constant in time, thereby mimicking a cosmological constant, but typically requires an extremely light scalar and therefore exhibits negligible clustering. In contrast, curvature-coupled models introduce a non-minimal interaction, which can be formulated either as a coupling to matter, as in coupled quintessence \cite{Amendola:1999er}, or as a coupling to curvature tensors, as in extended quintessence \cite{Perrotta:1999am}. These descriptions are physically equivalent and related by a conformal transformation.
The non-minimal coupling makes the scalar sector responsive to matter and metric perturbations, leading to a richer phenomenology than in minimally coupled quintessence. In particular, when the scalar sector is written in an effective fluid form, the curvature coupling terms can lead to an apparent crossing of the phantom divide, a behaviour realized in parts of the parameter space explored in this work.

Since curvature-coupled dark energy modifies both the background expansion and the evolution of perturbations, it is essential to understand how its parameters affect key cosmological observables. A central goal of this work is to identify regions of parameter space that remain compatible with current observations, and to characterize the corresponding signatures in the expansion history, the CMB, and the matter and metric potential spectra.

The background and linear perturbations of the curvature-coupled model are governed by coupled differential equations that cannot in general be solved analytically, so numerical Boltzmann solvers are required in order to compute the observable predictions. For this purpose we use \texttt{hi\_class} \cite{Zumalacarregui:2016pph,Bellini:2019syt}, an extension of \texttt{CLASS} \cite{Lesgourgues:2011re} tailored to Horndeski theories. Building on its internal structure, we implement the curvature-coupled dark energy model\footnote{\url{https://github.com/abdolalibanihashemi/hi_class_CCDE}} and solve its background and linear perturbation evolution in a fully consistent manner. 

The specific choice of coupling and potential adopted in this work is discussed in detail in Section~\ref{sec:model}. In particular, we consider a shifted quartic non-minimal coupling to Ricci scalar together with an inverse power-law scalar potential, a parametrization that provides a minimal and phenomenologically well-motivated realization of the curvature-coupled model. The inverse power-law potential belongs to the class of tracker quintessence potentials, allowing the scalar-field dynamics to approach an attractor solution over a wide range of initial conditions \cite{Zlatev:1998tr}, while the shift in the coupling enforces the correct present-day normalization of gravity and keeps the non-minimal interaction active throughout cosmic history without the need to introduce additional screening mechanisms. Other parametrizations, including thawing-gravity scenarios with effective phantom crossing, have also been considered in the literature, but generally face stronger tension with local gravity constraints unless screening effects are invoked \cite{Ye:2024ywg,Ye:2024zpk}. In the present paper we restrict our analysis to the background and linear regimes. A dedicated implementation of the curvature-coupled model in the relativistic $N$-body code \texttt{gevolution} \cite{Adamek:2015eda,Adamek:2016zes} will be pursued in future work. This is a natural direction, since several extensions of \texttt{gevolution} have already demonstrated its suitability for relativistic and modified gravity applications \cite{Hassani_2019, Hassani:2020rxd, Nouri-Zonoz:2025cul,Christiansen:2023tfy}.

The structure of this paper is as follows. In Section~\ref{sec:CCDE}, we introduce the theoretical framework of the curvature-coupled model and derive the corresponding equations of motion. In Section~\ref{sec:model}, we specify the form of the non-minimal coupling function and the scalar-field potential, motivated by stability criteria and local gravity constraints. In Section~\ref{sec:observables}, we present the predictions of the model for a set of key observables, including the expansion rate of the Universe, the CMB temperature angular power spectrum, the matter power spectrum, weak gravitational lensing, and the Integrated Sachs--Wolfe effect. Finally, in Section~\ref{sec:conclusions}, we summarize our results and discuss future perspectives.

\section{Theoretical Structure and the EFT Description}
\label{sec:CCDE}

The action for the curvature-coupled model in the Jordan frame reads as
\begin{equation}
    S = \int d^4x \sqrt{-g} \, \left\lbrace\frac{M_{\rm P}^2}{2} \left[1+f(\varphi)\right] R - \frac{1}{2} g^{\mu\nu}\partial_\mu\varphi\partial_\nu\varphi - V(\varphi) + \mathcal{L}_{m}[g_{\mu \nu}, \psi_i]\right\rbrace \,,\label{eq:action}
\end{equation}
where $g_{\mu\nu}$ and $g$ are the metric and its determinant, respectively, and $R$ is the Ricci scalar. The constant bare Planck mass is defined as $M_{\rm P}^2=(8\pi G)^{-1}$, where $G$ denotes the gravitational coupling that normalizes the Einstein--Hilbert term before the field-dependent modification $1+f(\varphi)$ is included. Its relation to the locally measured Newton constant is discussed in Sec.~\ref{local_tests}. The fields $\psi_i$ denote the standard matter fields, and $\mathcal{L}_m$ represents the Lagrangian density of the matter sector. We adopt units with $c=1$. The theory contains an additional scalar degree of freedom $\varphi$ compared to General Relativity (GR).
The action contains a kinetic term and a generic potential $V(\varphi)$, similar to standard quintessence, as well as a coupling function $f(\varphi)$, which introduces an interaction between the scalar field $\varphi$ and gravity. 
 Throughout this paper, we refer to this theory as the curvature-coupled model to emphasize the physical origin of the non-minimal interaction.
Varying the action \eqref{eq:action} with respect to the metric yields the modified Einstein equations in their covariant form,
\begin{align}
    & M_{\rm P}^2 [1 + f(\varphi)] G_{\mu\nu} - \nabla_\mu \varphi \nabla_\nu \varphi + g_{\mu\nu} \left( V(\varphi) + \frac{1}{2} g^{\rho\sigma} \nabla_\rho \varphi \nabla_\sigma \varphi \right) \nonumber\\ &
    - M_{\rm P}^2 f_{, \varphi}(\varphi) \left( \nabla_\mu \nabla_\nu \varphi - g_{\mu\nu} g^{\rho\sigma} \nabla_\rho \nabla_\sigma \varphi\right)   - M_{\rm P}^2 f_{, \varphi \varphi}(\varphi) \left( \nabla_\mu \varphi \nabla_\nu \varphi - g_{\mu\nu} g^{\rho\sigma} \nabla_\rho \varphi \nabla_\sigma \varphi\right) = T_{\mu\nu}\,,\label{eq:einstein}
\end{align}
where $T_{\mu \nu} := \frac{-2}{\sqrt{-g} } \frac{\delta(\sqrt{-g} \mathcal{L}_m)}{\delta g^{\mu \nu}}$ is the standard stress energy tensor and  $G_{\mu \nu} = R_{\mu \nu} - R g_{\mu \nu}/2$ denotes the Einstein tensor. A comma in the subscript denotes derivatives with respect to the scalar field,
such that $f_{,\varphi} \equiv  df/d\varphi$. Moreover, $\nabla_{\mu}$ represents the covariant derivative, which, for scalar quantities, simplifies to the partial derivative.
The second line of Eq.~\eqref{eq:einstein} encodes the contributions arising from the non-minimal coupling, proportional to $f_{,\varphi}$ and $f_{,\varphi\varphi}$. These terms originate from the
variation of the coupling term $f(\varphi)R$ in the action and describe
how the scalar field mediates modifications to the gravitational
dynamics.

By varying the action with respect to the scalar field, we derive the modified Klein-Gordon equation,
\begin{align}
    \nabla^\mu \nabla_\mu \varphi + \frac{M_{\rm P}^2}{2} R f_{, \varphi}(\varphi) - V_{, \varphi}(\varphi) = 0\,.\label{eq:klein-gordon-cov}
\end{align}

We note that the action (\ref{eq:action}) is well within the realm of Horndeski theories, which are specified through 
\begin{equation}
S[g_{\mu \nu},\varphi]=\int d^4x \sqrt{-g} \left[  \sum_{i=2}^5  \mathcal{L}_i[g_{\mu \nu},\varphi] + \mathcal{L}_{m}[g_{\mu \nu}, \psi_j] \right],
\label{eq:horndeski_action}
\end{equation}
where the four Lagrangian components, $\mathcal{L}_i$ are defined as
\begin{align}
\mathcal{L}_2 &= G_2 (\varphi, X)\, ,\\
\mathcal{L}_3 &= -G_3 (\varphi, X) \Box \varphi\, ,\\
\mathcal{L}_4 &=   G_4 (\varphi, X) R + G_{4X} (\varphi, X)\left[(\Box \varphi)^2-\varphi_{;\mu \nu}\varphi^{;\mu \nu}\right]\, ,\\
\mathcal{L}_5 &= G_5 (\varphi, X) G_{\mu \nu}\varphi^{;\mu \nu}-\frac{1}{6}G_{5X}(\varphi, X)\left[(\Box \varphi)^3-3(\Box \varphi)\varphi_{;\mu \nu}\varphi^{;\mu \nu}+2\varphi_{;\mu}^{\phantom{;\mu}\nu}\varphi_{;\nu}^{\phantom{;\nu}\alpha}\varphi_{;\alpha}^{\phantom{;\alpha}\mu}\right].
\end{align}
The functions $G_i(\varphi, X)$ are four arbitrary functions of the scalar field $\varphi$ and the kinetic term $X = -\frac{1}{2} \partial_\mu \varphi \partial^{\mu} \varphi$. Here, $\Box = \nabla_\mu \nabla^{\mu}$ denotes the d'Alembertian operator, and the subscript $X$ on $G_i$ indicates partial differentiation with respect to $X$, such that $G_{iX} \equiv \partial G_i / \partial X$. A semicolon denotes covariant differentiation; in particular, $\varphi_{;\mu}\equiv \nabla_\mu\varphi$ and $\varphi_{;\mu\nu}\equiv \nabla_\nu\nabla_\mu\varphi$.

By comparing the action in Eq.~\eqref{eq:action} with the Horndeski action in  Eq.~\eqref{eq:horndeski_action}, the $G_i$ functions for the theory can be expressed as,
\begin{align}
G_2(\varphi, X) &=  X - V(\varphi)\label{eq:G2} \,,\\
G_3(\varphi, X) &= 0\,,\\
G_4(\varphi, X) &=  \frac{M_{\rm P}^2}{2} \left[1 + f(\varphi) \right]\,,\\
G_5(\varphi, X) &= 0\,.
\label{eq:eq_lagrangian}
\end{align}

It is convenient to express  the curvature-coupled model in terms of what is called $\alpha_i$ functions. As it is shown in \citep{Bellini:2014fua}, these functions are enough and at the same time non-redundant to fully describe linear perturbations in any Horndeski theory.  This means that all the time dependent coefficients of linear operators can be generated out of these functions: 
\begin{itemize}
    \item $\alpha_{\rm K}$ (\emph{kineticity}): characterizes the kinetic energy of the scalar perturbations.
    \item $\alpha_{\rm B}$ (\emph{braiding}):  encodes the kinetic mixing between the scalar field and the metric.
    \item $\alpha_{\rm M}$ (\emph{Planck-mass run rate}): determines the running rate of the effective Planck mass.
    \item $\alpha_{\rm T}$ (\emph{tensor speed excess}): quantifies the excess in the speed of gravitational waves.
\end{itemize}
Given the $G_i$'s, one can calculate $\alpha_i$ functions following the recipe in \cite{Bellini:2014fua}. For the curvature-coupled model they read
\begin{align}
M_*^2(\varphi, X) &= M_{\rm P}^2 \left[ 1 + f(\varphi)\right]\,,\label{eq:Mstar_2}\\
\mathcal{H}M_*^2\alpha_{\rm M}(\varphi, X) &=  M_{\rm P}^2f_{, \varphi}(\varphi) {\varphi}'\,,\\
\mathcal{H}^2 M_*^2 \alpha_{\rm K}(\varphi, X) &=  \varphi '^{\, 2} \,,\\ 
\mathcal{H} M_*^2 \alpha_{\rm B}(\varphi, X) &= - M_{\rm P}^2 {\varphi}' f_{, \varphi}(\varphi)\,,\\
\alpha_{\rm T}(\varphi, X) &= 0\,, \label{alpha_i_functions}
\end{align}
where $\mathcal{H}=a'/a$ is the conformal Hubble function, and a prime over quantities means derivative with respect the conformal time, $\eta$. While in standard quintessence models \emph{braiding} and
\emph{Planck-mass running} are absent, they are generally non-zero in the
curvature-coupled model and contribute to both the background evolution and the linear
perturbations. However, the propagation speed of gravitational waves remains unity, in agreement with GR.

Finally, the theory can also be specified using the Effective Field Theory (EFT) of Dark Energy approach \cite{Gubitosi:2012hu}, which is based on writing an action for the perturbed Friedmann--Lemaître--Robertson--Walker (FLRW) metric that includes all terms invariant under time-dependent spatial diffeomorphisms up to quadratic order in perturbations. The action is constructed in the unitary gauge, where the constant-time slices coincide with hypersurfaces of uniform scalar field.
The non-vanishing EFT functions for this model, following the notation in \cite{Pogosian:2016pwr}, are given by:
\begin{align}
    \Omega (\eta) &= 1 + f(\varphi)\,,\label{eq:eft1}\\
    \Lambda (\eta) &= \frac{\varphi'^2}{2 a^2} - V(\varphi)\,,\label{eq:eft2}\\
    c(\eta) &= \frac{1}{2}\varphi'^2\label{eq:eft3}\,.
\end{align}
Note that in the curvature-coupled model, the theory consists of a quintessence Lagrangian supplemented by a non-minimal coupling to gravity. As a result, in the 3+1 decomposition in the unitary gauge, most terms, such as those proportional to $(\delta g^{00})^2$ or the perturbations of the extrinsic curvature, are absent, leading to the vanishing of most of the EFT coefficients in this theory.


\section{Model specification}\label{sec:model}

To obtain concrete predictions for the curvature-coupled model, the non-minimal coupling $f(\varphi)$ and the scalar potential $V(\varphi)$ must be specified. 
They should be chosen such that local tests of gravity and stability requirements are satisfied, after which their viability can be further tested against cosmological data. In this section, we first specify these functions and then discuss the corresponding viability
conditions. 

For the coupling appearing in the covariant action
\eqref{eq:action}, we
choose
\begin{align}
f(\varphi)
=
\alpha
\left(
\varphi^2-\varphi_{\rm today}^2
\right)^2 ,
\label{eq:DE_coupling2}
\end{align}
together with the inverse power-law potential
\begin{align}
V(\varphi)
=
\Lambda\,\varphi^{-\sigma}.
\label{eq:DE_potential2}
\end{align}
where $\alpha$ controls the strength of the non-minimal interaction,
$\sigma$ determines the steepness of the potential, and $\Lambda$ sets its
overall amplitude. Here $\sigma$ is dimensionless, while $\alpha$ has mass dimension $-4$
and $\Lambda$ has mass dimension $4+\sigma$. This is because the scalar field $\varphi$ appearing in the action
\eqref{eq:action} is canonically normalized and has mass dimension one.
In the numerical implementation, though, we work in Planck units, $M_{\rm P}=1$,
following the \texttt{hi\_class} convention, so the dimensionful quantities quoted
in the numerical results are therefore understood in the corresponding
powers of $M_{\rm P}$ (see Appendix~\ref{app:hiclass}).
The quantity $\varphi_{\rm today}$ denotes the present value of the background
scalar field, but in Eq.~\eqref{eq:DE_coupling2} it enters as a constant shift
rather than as an independent parameter specified in advance. This shift is
fixed by requiring the non-minimal coupling to vanish at the present time, so that the effective Planck mass recovers its standard
normalization today. In practice, \texttt{hi\_class} enforces this condition by
adjusting $\varphi_{\rm today}$ until the evolved background satisfies
$M_*^2(z=0)=M_{\rm P}^2$.

The inverse power-law potential is well motivated in the dark energy literature, since it admits tracker solutions for $\sigma>0$ and can drive late-time acceleration without introducing an explicit cosmological constant \citep{Zlatev:1998tr}. 
The shifted coupling in Eq.~\eqref{eq:DE_coupling2}, in turn, allows the scalar field to modify gravity cosmologically while preserving the correct local normalization of Newton's constant today.

The amplitude $\Lambda$ is not treated as an independent parameter, but is fixed by the closure relation
\begin{equation}
1 - \sum_{i \neq \varphi} \Omega_i = \Omega_\varphi,
\end{equation}
so that the scalar field accounts for the observed dark energy density today. In practice, the shooting algorithm fixes both the potential normalization $\Lambda$, and the shift $\varphi_{\rm today}$,
imposing simultaneously the closure relation and the present-day normalization
$M_*^2(z=0)=M_{\rm P}^2$.

Relative to $\Lambda$CDM, the model therefore introduces two free parameters $\{\alpha,\sigma\}$, together with the initial conditions
$\varphi_{\rm ini}$ and $\varphi'_{\rm ini}$. For $\sigma>0$, the tracker behaviour of the inverse power-law potential makes the late-time dynamics largely insensitive to the precise initial conditions, provided they lie within the basin of attraction.
  The late-time phenomenology is therefore effectively controlled by $\{\alpha,\sigma\}$.

The $\Lambda$CDM limit is recovered for $\alpha=0$ and $\sigma=0$, in which case $V(\varphi)=\Lambda$ acts as a cosmological constant and the scalar field remains constant\footnote{For numerical stability, it is preferable to keep $\varphi$ constant rather than setting $\varphi=0$, since $\sigma>0$ would otherwise lead to a singular denominator.}. This limit provides a useful consistency check of the implementation.
A distinctive feature of this parametrization is that the shift by $\varphi_{\rm today}$ allows $f(\varphi)$ to remain non-zero even when $\varphi$ is small, keeping the non-minimal coupling active throughout cosmic
history. The model can therefore affect both the pre-recombination expansion history, potentially shifting the CMB-inferred value of $H_0$, and the late-time cosmological dynamics.


\subsection{Stability requirements}
\label{subsec:stability}
A viable realization of the curvature-coupled model must remain free of pathological
instabilities throughout cosmic history. Since the curvature-coupled model belongs to the Horndeski class, Ostrogradsky instabilities are absent by construction, as the equations of motion remain second order \cite{Ostrogradsky:1850fid}. The non-trivial viability conditions therefore reduce to the absence of ghost and gradient instabilities in the scalar and tensor sectors.

In unitary gauge, where the scalar-field fluctuation is absorbed into the time slicing, the dynamics of the propagating scalar and tensor modes is encoded in the quadratic action
\begin{equation}\label{secorderaction}
S_2 = \int d\eta\,d^3x\,a^2 \left\{
Q_{\rm s}\left[\zeta'^2 - c_{\rm s}^2 (\partial_i \zeta)^2 \right]
+
Q_{\rm T}\left[h_{ij}'^2 - c_{\rm T}^2 (\partial_k h_{ij})^2 \right]
\right\},
\end{equation}
where $\zeta$ denotes the scalar propagating degree of freedom in unitary gauge, and $h_{ij}$ are the elemnts of the transverse-traceless tensor perturbations. The coefficients $Q_{\rm s}$ and $Q_{\rm T}$ are the kinetic normalizations of the scalar and tensor sectors, since they multiply the time-derivative terms in the quadratic action. Their positivity guarantees the absence of ghosts, while $c_{\rm s}^2$ and $c_{\rm T}^2$ control the gradient terms and must be positive to avoid gradient instabilities. In Horndeski theories these quantities can be written in terms of the background quantities and the EFT $\alpha$-functions introduced above in Eqs.~\eqref{eq:Mstar_2}--\eqref{alpha_i_functions} \cite{Bellini:2014fua}. For the curvature-coupled model, one finds 
\begin{align}
& Q_{\rm s} = \frac{2 M_{\rm P}^2\left[1+f(\varphi)\right]\left(\alpha_{\rm K}+\frac{3}{2}\alpha_{\rm B}^2\right)}{(2-\alpha_{\rm B})^2}, \\
& c_{\rm s}^2 = -\frac{(2-\alpha_{\rm B})\left[\mathcal{H}'-\mathcal{H}^2\left(1+\frac{1}{2}\alpha_{\rm B}+\alpha_{\rm M}\right)\right]-\mathcal{H}{\alpha}_{\rm B}'+\frac{a^2}{M_{\rm P}^2(1+f(\varphi))}(\rho_m+P_m)}{\mathcal{H}^2\left(\alpha_{\rm K}+\frac{3}{2}\alpha_{\rm B}^2\right)}, \\
& Q_{\rm T} = \frac{M_{\rm P}^2}{8}\left[1+f(\varphi)\right], \\
& c_{\rm T}^2 = 1 + \alpha_{\rm T},
\end{align}
where $\rho_m$ and $P_m$ denote the total background energy density and pressure of the standard matter sector, i.e. the sum over all minimally coupled fluid components.
Stability requires
\be
Q_{\rm s} > 0, \qquad Q_{\rm T} > 0 \quad \text{(no ghosts)}, \nonumber
\ee
and
\be
c_{\rm s}^2 > 0, \qquad c_{\rm T}^2 > 0 \quad \text{(no gradient instabilities)}. \nonumber
\ee
Vector perturbations do not introduce additional conditions \cite{DeFelice:2011bh}.

For the curvature-coupled model, $\alpha_{\rm T}=0$ \eqref{alpha_i_functions}, implying $c_{\rm T}^2=1$. The tensor sector is therefore automatically free of gradient instabilities and consistent with the GW170817/GRB170817A bound on the speed of gravitational waves \cite{PhysRevLett.119.161101, Abbott2017:GRGW}. Moreover,
\begin{equation}
Q_{\rm T} = \frac{M_{\rm P}^2}{8}\left[1+f(\varphi)\right],
\end{equation}
so the absence of tensor ghosts requires
\begin{equation}
1 + f(\varphi) > 0.
\end{equation}
This is equivalent to the positivity of the effective Planck mass squared,
\begin{equation}
M_*^2 = M_{\rm P}^2\left[1+f(\varphi)\right].
\end{equation}

For the scalar sector,
\begin{equation}
\alpha_{\rm K}
=
\frac{{\varphi}'^{\,2}}
{\mathcal{H}^2 M_{\rm P}^2[1+f(\varphi)]}
\end{equation}
is positive for $\varphi'\neq0$ provided $1+f(\varphi)>0$, and hence
$Q_{\rm s}>0$ away from turning points of the background scalar field ($\varphi'=0$).

It is worth noting that at isolated times where $\varphi'=0$, both $\alpha_{\rm K}$ and
$\alpha_{\rm B}$ vanish, and the unitary-gauge description becomes
degenerate because the scalar field cannot be used to define the time
slicing. The resulting $Q_{\rm s}=0$ at the turning point therefore does
not by itself signal a physical instability, but rather the breakdown of the effective field theory construction within this formalism. In particular, the
scalar-field perturbation $\delta\varphi$ and the corresponding covariant
perturbation equations remain regular across the crossing, as discussed
in Sec.~\ref{sec:linear_pert} and Appendix~\ref{app:pert_eqs}.

Away from these isolated turning points, using the background equations
of motion derived in Sec.~\ref{BG_analysis}, namely
Eqs.~\eqref{eq:friedmann1}, \eqref{eq:friedmann2}, and
\eqref{eq:klein-gordon}, together with the definitions of the
$\alpha_i$ in Eqs.~\eqref{eq:Mstar_2}--\eqref{alpha_i_functions},
one finds that $c_{\rm s}^2=1$ identically. Scalar perturbations therefore
propagate luminally and are free of gradient instabilities wherever the
unitary-gauge description is well defined.

The only physical stability requirement is therefore
\begin{equation}
f(\varphi)>-1,
\label{eq:stability_condition}
\end{equation}
which ensures a positive effective Planck mass throughout the evolution,
with the unitary-gauge scalar stability conditions understood away from
the isolated points where $\varphi'=0$.


\subsection{Consistency with local gravity constraints}
 \label{local_tests}
Although there is no fundamental reason to forbid a cosmological evolution of
the effective Planck mass, such an evolution can affect quantities whose
present-day values are tightly constrained by laboratory and Solar-System
experiments. Since the curvature-coupled model considered here is specified
by a covariant scalar--tensor action, rather than only through a
phenomenological parametrization, the same theory must also be consistent
with local tests of gravity.

To connect the cosmological evolution of the scalar field with its local
weak-field behaviour, we decompose the field into its homogeneous
cosmological background and a local perturbation,
\begin{equation}
\varphi(t,\mathbf{x})
=
\bar{\varphi}(t)
+
\delta\varphi(t,\mathbf{x}),
\label{eq:local_phi_decomposition}
\end{equation}
where $\bar{\varphi}(t_0)\equiv\varphi_{\rm today}$ at the present epoch.
For a local gravitational system embedded in the cosmological background,
we impose the asymptotic boundary condition
$\delta\varphi\rightarrow0$ away from the local source, such that
$\varphi\rightarrow\varphi_{\rm today}$ at the present epoch.

The response of the scalar field to the local matter distribution follows
from the Klein--Gordon equation, Eq.~\eqref{eq:klein-gordon-cov}.
Writing the local curvature as $R(\mathbf{x})=\bar R+\delta R(\mathbf{x})$,
where $\delta R$ is generated by the local matter distribution, we linearize
the scalar equation in $\delta\varphi$ around
$\varphi_{\rm today}$, while retaining the local curvature
$R(\mathbf{x})$. In the quasistatic weak-field regime, and retaining the dominant local
curvature dependence, this gives schematically
\begin{equation}
\left[\nabla^2-m_{\rm eff}^2(\mathbf{x})\right]\delta\varphi
\simeq
-\frac{M_{\rm P}^2}{2}
f_{,\varphi}(\varphi_{\rm today})\,\delta R(\mathbf{x}),
\label{eq:local_scalar_weakfield}
\end{equation}
where the effective mass is
$m_{\rm eff}^2(\mathbf{x})
=
V_{,\varphi\varphi}(\varphi_{\rm today})
-
{M_{\rm P}^2 R(\mathbf{x})}
f_{,\varphi\varphi}(\varphi_{\rm today})/2$. Compared with the full linearized Klein--Gordon equation in
Eq.~\ref{eq:kg_pert_eq_app}, we neglect here terms proportional to the
weak metric potentials, such as $V_{,\varphi}\Psi$, which are subleading
compared with the local curvature contribution in the weak-field regime
considered here.

For the shifted coupling in Eq.~\eqref{eq:DE_coupling2},
$f(\varphi_{\rm today})=f_{,\varphi}(\varphi_{\rm today})=0$.
With the source term vanishing, Eq.~\eqref{eq:local_scalar_weakfield}
reduces to the homogeneous equation
\begin{equation}
\left[\nabla^2-m_{\rm eff}^2(\mathbf{x})\right]\delta\varphi=0.
\end{equation}
Multiplying by $\delta\varphi$, integrating over space, and using
$\delta\varphi\rightarrow0$ at spatial infinity gives
\begin{equation}
\int {\rm d}^3x\,
\left[
|\boldsymbol{\nabla}\delta\varphi|^2
+
m_{\rm eff}^2(\mathbf{x})\,\delta\varphi^2
\right]
=0.
\label{eq:local_stability_integral}
\end{equation}
Therefore, if $m_{\rm eff}^2(\mathbf{x})\geq0$ throughout the local
environment, the only solution satisfying the asymptotic boundary
condition is $\delta\varphi=0$.

For the coupling in Eq.~\eqref{eq:DE_coupling2},
\begin{equation}
f_{,\varphi\varphi}(\varphi_{\rm today})
=
8\alpha\varphi_{\rm today}^2,
\end{equation}
and hence
\begin{equation}
m_{\rm eff}^2(\mathbf{x})
=
V_{,\varphi\varphi}(\varphi_{\rm today})
-
4\alpha M_{\rm P}^2
\varphi_{\rm today}^2 R(\mathbf{x}).
\label{eq:local_meff_explicit}
\end{equation}
Since $V_{,\varphi\varphi}>0$ for $\Lambda>0$ and $\sigma>0$, the
curvature contribution is positive for $\alpha<0$ in regions with
$R\geq0$, so that $m_{\rm eff}^2>0$ and the linear weak-field branch
$\delta\varphi=0$ is stable under these conditions.
For $\alpha>0$, the curvature term instead lowers the effective mass and
$m_{\rm eff}^2$ can become negative in sufficiently large-curvature
regions. A negative local value of $m_{\rm eff}^2$ does not, however, automatically
imply an instability. This can already be seen from
Eq.~\eqref{eq:local_scalar_weakfield}, that for a scalar perturbation varying
over a characteristic length scale $L$, the gradient term scales as
$\nabla^2\delta\varphi\sim\delta\varphi/L^2$, whereas the effective-mass
term scales as $m_{\rm eff}^2\delta\varphi$. The negative mass contribution
can therefore compete with the stabilizing gradient term only when
\begin{equation}
|m_{\rm eff}^2|L^2\gtrsim\mathcal{O}(1).
\end{equation}
The same competition between the spatial-gradient and mass terms is
explicit in the full linearized Klein--Gordon equation,
Eq.~\eqref{eq:kg_pert_eq_app}.

For comparison, if $f_{,\varphi}(\varphi_{\rm today})$ were nonzero and the
effective mass could be treated as approximately constant outside the source,
the exterior solution would have the Yukawa form
\begin{equation}
\delta\varphi(r)
\propto
f_{,\varphi}(\varphi_{\rm today})
\frac{e^{-m_{\rm eff}r}}{r}.
\end{equation}
For a light scalar satisfying $m_{\rm eff}r\ll1$ on Solar-System scales,
such a scalar-mediated interaction would be long ranged and therefore
strongly constrained by local tests. In the present model, however, the
amplitude of this linear profile vanishes because
$f_{,\varphi}(\varphi_{\rm today})=0$.

Assuming that the linear weak-field solution $\delta\varphi=0$ is valid and stable,
we evaluate the local-gravity observables below at
$\varphi=\varphi_{\rm today}$. Possible environmental and non-linear
departures from this solution are discussed afterwards.

\begin{itemize}
\item Constraint on $G_{\rm eff}^{\rm today}$.
Cavendish-type experiments measure the gravitational force, $F_{\rm g}$,
between two test bodies of masses $m_1$ and $m_2$, separated by a distance
$r$, and thereby determine the locally measured gravitational coupling,
conventionally denoted by Newton's constant, $G_{\rm N}$ \cite{1252701}.
Hence, at the present epoch one requires
$G_{\rm eff}^{\rm today}=G_{\rm N}$.
Following \cite{Esposito-Farese:2000pbo}, in curvature-coupled
scalar--tensor theories the effective gravitational coupling measured in
local experiments is
\begin{equation}
G_{\rm eff}
=
\frac{G}{1+f(\varphi)}
\left[1+\varepsilon^2(\varphi)\right],
\qquad
\varepsilon^2(\varphi)
\equiv
\frac{M_{\rm P}^2 f_{,\varphi}(\varphi)^2}
{2+2f(\varphi)+3M_{\rm P}^2 f_{,\varphi}(\varphi)^2},
\label{eq:Geff}
\end{equation}
where $G$ is the bare gravitational coupling entering the action in
Eq.~\eqref{eq:action}. For the parametrization in
Eq.~\eqref{eq:DE_coupling2},
$f(\varphi_{\rm today})=f_{,\varphi}(\varphi_{\rm today})=0$, and therefore
$\varepsilon(\varphi_{\rm today})=0$. Eq.~\eqref{eq:Geff} then gives $G_{\rm eff}^{\rm today}=G$.
Since the present-day locally measured coupling is $G_{\rm N}$, it follows
that $G=G_{\rm N}$ for our model.

\item Constraint on $\gamma_{\rm PPN}$. 
The parameter $\gamma_{\rm PPN}$ characterizes the ratio of the two
metric potentials in the local weak-field limit. Measurements of the
Shapiro time delay, most notably from the Cassini mission
\cite{Bertotti:2003rm}, constrain it to
$\gamma_{\rm PPN}=1+(2.1\pm2.3)\times10^{-5}$.
In our model, it is given by \citep{Will:2014kxa}
\begin{equation}
\gamma_{\rm PPN}
=
1-\frac{2\varepsilon^2(\varphi)}
{1+\varepsilon^2(\varphi)}.
\label{eq:gamma}
\end{equation}
Since $\varepsilon(\varphi_{\rm today})=0$, the GR value
$\gamma_{\rm PPN}=1$ is recovered today, consistently with the observational bound.

\item Constraint on $\beta_{\rm PPN}$. This parameter quantifies the nonlinearity of gravity and is constrained by observations of the perihelion precession of Solar System planets. Solar-System measurements constrain  $\beta_{\rm PPN}-1=(-4.1\pm7.8)\times10^{-5}$ \cite{Fienga:2011qh}. In our model, the deviation scales as
$\beta_{\rm PPN}-1\propto\varepsilon^2(\varphi)$
\citep{Uzan:2010pm}. Since
$\varepsilon(\varphi_{\rm today})=0$, the deviation vanishes at the present
epoch and the GR value $\beta_{\rm PPN}=1$ is recovered.
\item Constraint on $\dot{G}_{\rm eff}/G_{\rm eff}$. This quantity can be constrained, for example, by decades of precise lunar laser ranging measurements, which place an upper bound on the present-day fractional time variation of Newton's constant, 
\be
\left|\frac{\dot G_{\rm eff}}{G_{\rm eff}}\right|_{\rm today} \lesssim 10^{-11}\,{\rm yr}^{-1}, \nonumber
\ee
where the derivative is taken with respect to physical time \cite{Gillies_1997}. Taking the derivative with respect to physical time of Eq.~\eqref{eq:Geff} and using $\dot{G}_{\rm eff}=(dG_{\rm eff}/d\varphi)\dot\varphi$, we obtain
\begin{equation}
\dot G_{\rm eff}
=
-\frac{G_{\rm N}\dot\varphi}{\left(1+f(\varphi)\right)^2}
\left[ 
\bigl(1+ \varepsilon^2(\varphi)\bigr)\,f_{,\varphi}(\varphi)
 - 2\bigl(1+ f(\varphi)\bigr)\, \varepsilon(\varphi) \varepsilon_{,\varphi}(\varphi)
\right].
\end{equation}
Equivalently, the fractional variation can be written as
\begin{equation}
\frac{\dot G_{\rm eff}}{G_{\rm eff}}
=
-\dot\varphi
\left[
\frac{f_{,\varphi}(\varphi)}{1+f(\varphi)}
-
\frac{2 \varepsilon(\varphi) \varepsilon_{,\varphi}(\varphi)}{1+ \varepsilon^2(\varphi)}
\right].
\label{eq:Gdot_over_G}
\end{equation}
Evaluating Eq.~\eqref{eq:Gdot_over_G} at the present time, and using the fact that in our parameterization
$f(\varphi_{\rm today})=f_{,\varphi}(\varphi_{\rm today})=\varepsilon(\varphi_{\rm today})=0$, we obtain
\be
\left.\frac{\dot G_{\rm eff}}{G_{\rm eff}}\right|_{\rm today}=0. \nonumber
\ee
Hence, the model automatically satisfies the present observational bound on the time variation of the effective gravitational coupling.
\end{itemize}
The weak-field analysis above shows that
$\varphi=\varphi_{\rm today}$ is a solution of the linearized local scalar
equation, since $f_{,\varphi}(\varphi_{\rm today})=0$. This does not,
however, determine the full non-linear scalar profile in a realistic
environment. If the local solution departs from
$\varphi_{\rm today}$, then $f_{,\varphi}(\varphi)$ becomes nonzero and the
curvature-dependent term $M_{\rm P}^2 R f_{,\varphi}(\varphi)/2$ can modify the
local scalar dynamics. Since $R$ depends on the local matter environment, the curvature coupling
induces an environment-dependent effective potential,
$ V_{\rm eff}(\varphi,R)
=
V(\varphi)
-
\frac{M_{\rm P}^2}{2}R\,f(\varphi),
$
as follows directly from Eq.~\eqref{eq:klein-gordon-cov}. While the
linearized behaviour around $\varphi_{\rm today}$ depends on the sign of
$\alpha$ as discussed above, the full non-linear response is determined by
the complete shape of $V_{\rm eff}(\varphi,R)$. In sufficiently dense regions, such a potential may develop a local minimum with a large effective scalar mass, thereby shortening the scalar Compton
wavelength and suppressing the scalar-mediated force. This density-dependent
suppression is characteristic of a chameleon-like screening mechanism and can
help recover the GR limit locally. Determining whether this occurs for the parameter space considered here
requires solving the non-linear scalar-field equation for realistic matter
distributions, potentially within a dedicated $N$-body implementation.

By contrast, the action in Eq.~\eqref{eq:action} does not contain the derivative
operators required for Vainshtein or K-mouflage screening. Vainshtein screening requires derivative self-interactions beyond the canonical
kinetic term, involving non-linear combinations of second derivatives of the
scalar field, $\nabla_\mu\nabla_\nu\varphi$, as in Galileon operators. K-mouflage screening instead
relies on non-linear dependence on the first-derivative kinetic term, $X$. Both derivative-screening structures are absent here, since the action
contains only a canonical kinetic term, a potential $V(\varphi)$, and a
non-minimal coupling depending only on $\varphi$.
See~\citep{Sirera:2026klo,Joyce:2014kja,Clifton:2011jh,Koyama:2015vza}
for reviews of screening mechanisms in modified gravity theories.


\paragraph{Combined theoretical and observational constraints.}

The viable parameter space is therefore restricted by the stability condition
$f(\varphi)>-1$. In addition, $\Lambda$ is fixed by the closure relation, while $\varphi_{\rm today}$ is fixed by imposing the present-day gravitational
normalization $M_*^2(z=0)=M_{\rm P}^2$.
The cosmological phenomenology is therefore effectively
controlled by the remaining parameters $\{\alpha,\sigma\}$.

\section{Cosmological implications}
\label{sec:observables}
The curvature-coupled model affects cosmological observables through three main channels: they modify the background expansion history, change the growth of matter perturbations, and alter the evolution of the metric potentials. These effects leave signatures in distance measures, the CMB, large-scale structure, galaxy clustering, weak lensing, and ISW observables. In this section, we present the corresponding predictions computed with \texttt{hi\_class}, starting from the background dynamics and then turning to the linear perturbation sector. 

Unless otherwise stated, all spectra and background quantities shown in the
following discussion are computed at fixed baseline cosmological parameters
consistent with the Planck~2018 TT+lowE $\Lambda$CDM best-fit values
\citep{Planck:2018vyg}.
\[
H_0 = 66.88\,{\rm km\,s^{-1}\,Mpc^{-1}},\qquad
\omega_b = 0.02212,\qquad
\omega_{\rm cdm}=0.1206,
\]
\[
A_s = 2.092\times 10^{-9},\qquad
n_s = 0.9626,\qquad
\tau_{\rm reio}=0.0522.
\]
We include one massive neutrino species with
\(m_{\rm ncdm}=0.06\,{\rm eV}\), \(N_{\rm ncdm}=1\), and
\(N_{\rm ur}=2.046\). 

To illustrate the dependence on the model parameters, we consider
representative combinations of $\sigma$ and $\alpha$, varying both the
potential slope and the strength and sign of the non-minimal coupling.
For the $(\sigma,\alpha)$ values considered in the
background and perturbation analyses below, the case
$(\sigma,\alpha)=(0.001,0.001)$ is numerically indistinguishable from
$\Lambda$CDM over the ranges shown. We therefore omit a separate
$\Lambda$CDM curve where the two overlap, while all relative differences
are computed with respect to the $\Lambda$CDM prediction.

\subsection{Background analysis \label{BG_analysis}}
We assume that at large enough scales the Universe is homogeneous,
isotropic, and spatially flat, and is therefore described by the flat FLRW
metric,
\begin{equation}
    ds^2 = a^2(\eta) \left( - d \eta^2 +  \delta_{ij} dx^i dx^j\right)\,.
\end{equation}
Additionally, we assume that the standard matter sector (baryonic matter, cold dark matter, and relativistic species) is described by a perfect fluid stress-energy tensor
\begin{equation}
    T^{\mu\nu} = \left(\rho + P\right) u^\mu u^\nu + P g^{\mu\nu} \,.
\end{equation}
Here, $u^\mu = dx^\mu/d\tau$ represents the fluid velocity with $\tau$ being the proper time, $\rho$ is the fluid's rest frame energy density, and $P$ is its pressure.
With the background FLRW metric, the modified Einstein equations \eqref{eq:einstein} will reduce to the $\eta$-$\eta$ and $i$-$i$ components, which are given as follows:
\begin{align}
    \mathcal{H}^2 &= \frac{a^2}{3 M_{\rm P}^2\left[1+f(\varphi)\right]}\left(\rho + \frac{\varphi'^2}{2 a^2} + V(\varphi) - \frac{3M_{\rm P}^2 \mathcal{H}\varphi'f_{,\varphi}(\varphi)}{a^2} \right)\,,\label{eq:friedmann1} \\
    \mathcal{H}^2+2\mathcal{H}' &= - \frac{a^2}{M_{\rm P}^2 \left[1+f(\varphi)\right]}\left(P + \frac{\varphi'^2}{2 a^2}- V(\varphi) + \frac{M_{\rm P}^2}{a^2} (\mathcal{H}\varphi' + \varphi'')f_{,\varphi}(\varphi) + \frac{M_{\rm P}^2 \varphi'^2 f_{,\varphi\varphi}(\varphi)}{a^2} \right)\,,\label{eq:friedmann2}
\end{align}
while the modified Klein--Gordon equation \eqref{eq:klein-gordon-cov} becomes
\begin{equation}
    \varphi'' + 2 \mathcal{H} \varphi' + a^2 V_{,\varphi}(\varphi) - 3 M_{\rm P}^2 (\mathcal{H}^2+\mathcal{H}') f_{,\varphi}(\varphi) = 0\,.\label{eq:klein-gordon}
\end{equation}
The $^\prime$ denotes the derivative with respect to conformal time $\eta$, and the conformal Hubble parameter is defined as $\mathcal{H} \equiv a'/a$. These equations govern the dynamics of scale factor, the background scalar field, and the background densities of the matter species. It is also useful to characterize the field evolution per e-fold,
\begin{equation}
\frac{d\varphi}{d\ln a}=\frac{\varphi'}{\mathcal{H}},
\end{equation}
which directly measures how rapidly the scalar evolves relative to the cosmological expansion.
Revisiting the covariant Klein--Gordon equation \eqref{eq:klein-gordon-cov}, or its spatially coarse grained one,  Eq.~\eqref{eq:klein-gordon}, makes it clear that the scalar-field dynamics is driven not only by the bare potential $V(\varphi)$ 
but also by the curvature coupling through the term $f_{,\varphi}(\varphi) R$. 
It is therefore convenient to define an effective potential,
\begin{equation}
    V_{\rm eff}(\varphi) \equiv V(\varphi) - \frac{M_{\rm P}^2}{2}\, R\, f(\varphi),
\end{equation}
so that the Klein--Gordon equation may be written schematically as 
$\nabla^\mu\nabla_\mu\varphi - V_{\rm eff,\,\varphi} = 0$.  
The evolution of the scalar field is thus governed by the combined effect of the 
bare potential slope and the curvature-induced contribution. 
Different choices of the coupling parameter $\alpha$ and the potential exponent~$\sigma$ in Eqs.~\eqref{eq:DE_coupling2}--\eqref{eq:DE_potential2} change the shape and redshift dependence of $V_{\rm eff}(\varphi)$, altering the 
effective force acting on the field and thereby producing qualitatively distinct 
background evolutions.

On the other hand, from the modified Friedmann equations Eqs.~\eqref{eq:friedmann1}--\eqref{eq:friedmann2} we note that one can define the \emph{effective} energy density and pressure of the scalar field such that the standard GR form of the Friedmann equations holds,
\begin{align}
    \mathcal{H}^2 &= \frac{a^2}{3 M_{\rm P}^2}\left(\rho_m +\rho_\varphi\right)\,,\label{eq:friedmann3} \\
    \mathcal{H}^2+2\mathcal{H}' &= - \frac{a^2}{M_{\rm P}^2 }\left(P_m +P_\varphi \right)\,,\label{eq:friedmann4}
\end{align}
with $\rho_\varphi$ and $P_\varphi$ being,
\begin{align}
    \rho_\varphi &= \frac{\varphi'^2}{2a^2}+V(\varphi)-\frac{3M_{\rm P}^2\mathcal{H}\varphi'f_{,\varphi}(\varphi)}{a^2}-\frac{3M_{\rm P}^2f(\varphi)\mathcal{H}^2}{a^2}\,,\label{density}\\
    P_\varphi &=\frac{\varphi'^2}{2 a^2}- V(\varphi) + \frac{M_{\rm P}^2}{a^2} (\mathcal{H}\varphi' + \varphi'')f_{,\varphi}(\varphi) + \frac{M_{\rm P}^2 {\varphi'}^2 f_{,\varphi\varphi}(\varphi)}{a^2}+\frac{M_{\rm P}^2}{a^2}f(\varphi)\left(\mathcal{H}^2+2{\mathcal{H}'}\right)\label{pressure}\,.
\end{align}

It is important to note that the definitions of $\rho_\varphi$ and $P_\varphi$ in 
Eqs.~(\ref{density})--(\ref{pressure}) may appear non-standard, as they explicitly 
contain the quantities $\mathcal H^2$ and ${\mathcal H'}$. In some situations 
these definitions can even lead to a negative effective energy density for the 
scalar field. Nevertheless, this formulation is convenient because it allows the 
Friedmann equations to retain their standard GR form without redefining the 
Planck mass, while providing a consistent bookkeeping of the contributions from 
the scalar field and its non-minimal coupling to gravity. In addition, with these 
definitions the scalar sector obeys the usual continuity equation,
\begin{equation}
{\rho}'_\varphi + 3\mathcal H(\rho_\varphi + P_\varphi)=0 .
\end{equation}

At early times the scalar field is effectively frozen by the large Hubble friction term in
the Klein--Gordon equation, with
$|d\varphi/d\ln a|=|\varphi'|/\mathcal{H}\ll M_{\rm P}$,
so that the kinetic contribution to $\rho_\varphi$ in Eq.~(4.9) is negligible. In this regime the effective scalar
energy density reduces approximately to
\begin{equation}
    \rho_\varphi \simeq V(\varphi)
    -\frac{3M_{\rm P}^2 f(\varphi)\mathcal H^2}{a^2}.
\end{equation}
The relative importance of these two terms depends on the initial
value of the field and on the model parameters. In particular, since
the potential is of inverse power-law form, a small value of
$\varphi_{\rm ini}$ can enhance $V(\varphi_{\rm ini})$. However, the
initial conditions considered here are chosen such that the early
Universe remains radiation dominated, i.e. the scalar contribution is
subdominant compared with the radiation density in Eq.~\eqref{eq:friedmann1}. During radiation domination one has, to leading order,
\begin{equation}
    \frac{3M_{\rm P}^2\mathcal H^2}{a^2} \simeq \rho_r ,
\end{equation}
where $\rho_r$ denotes the background radiation energy density. This relation holds up to corrections from the non-minimal coupling and from the subdominant matter and scalar-field components. Therefore the coupling
contribution to the effective scalar density scales approximately as
$-f(\varphi)\rho_r$. For the parameter choices and initial conditions
used in our numerical analysis, this term dominates over the bare
potential contribution at sufficiently high redshift, so that
\begin{equation}
    \rho_\varphi \simeq
    -\frac{3M_{\rm P}^2 f(\varphi)\mathcal H^2}{a^2}.
\end{equation}
Since the shifted coupling satisfies
$f(\varphi_{\rm ini})\simeq \alpha \varphi_{\rm today}^4$ when
$\varphi_{\rm ini}\ll \varphi_{\rm today}$, the early-time effective
scalar density is negative for $\alpha>0$ and positive for
$\alpha<0$, within this curvature-dominated regime. This contribution
should not be interpreted as the energy density of an independent
physical fluid; rather, it arises from rewriting the modified
 equations in the standard Friedmann form.

The background evolution of the scalar field and the fractional energy densities is illustrated in Fig.~1.
The left panel shows $\varphi(z)$, while the middle panel shows
$|d\varphi/d\ln a|=|\varphi'|/\mathcal{H}$, with solid and dashed curves indicating positive and negative field evolution, respectively.
At high redshift this quantity is strongly suppressed, showing that the field is effectively frozen by Hubble friction. As the expansion rate decreases at 
later times, the field gradually begins to evolve and roll down its effective
potential. The details of this transition depend on both the steepness 
of the potential $V(\varphi)\propto\varphi^{-\sigma}$ and the strength 
of the non-minimal coupling parameter $\alpha$.

\begin{figure}[h]
\centering
\includegraphics[scale=0.262]{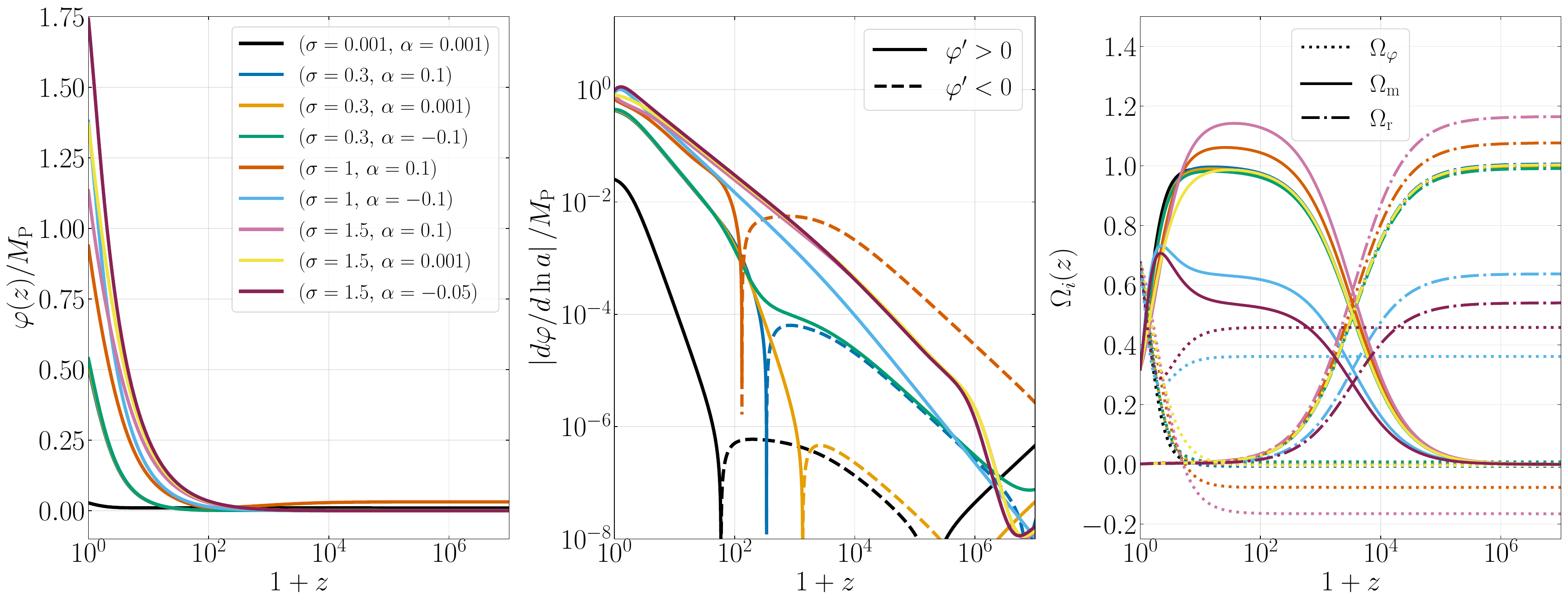}
\caption{Background evolution of the scalar field and fractional energy densities for 
several representative parameter choices. 
Left: Planck-normalized scalar-field amplitude $\varphi(z)/M_{\rm P}$. 
Middle: magnitude of the scalar-field variation per e-fold,
$|d\varphi/d\ln a|/M_{\rm P}=|\varphi'|/(\mathcal{H}M_{\rm P})$,
with dashed curves indicating negative values. 
Right: fractional energy densities $\Omega_i(z)$ of the scalar field, matter, 
and radiation components. 
At early times the scalar field is frozen by Hubble friction, while the 
curvature--coupling term can lead to a negative effective scalar density for 
$\alpha>0$.
}
\label{fig:Omega_i}
\end{figure}

The right panel of Fig.~\ref{fig:Omega_i} shows how this dynamics affects the 
background energy budget. In particular, for positive values of the coupling 
parameter $\alpha$, the effective scalar energy density becomes negative at early 
times, as expected from the last term in Eq.~(\ref{density}). Because the total 
energy density must remain equal to the critical density, the fractional densities 
of the other components temporarily exceed unity to compensate for this negative 
contribution.

\begin{figure}[h]
\centering
\includegraphics[scale=0.262]{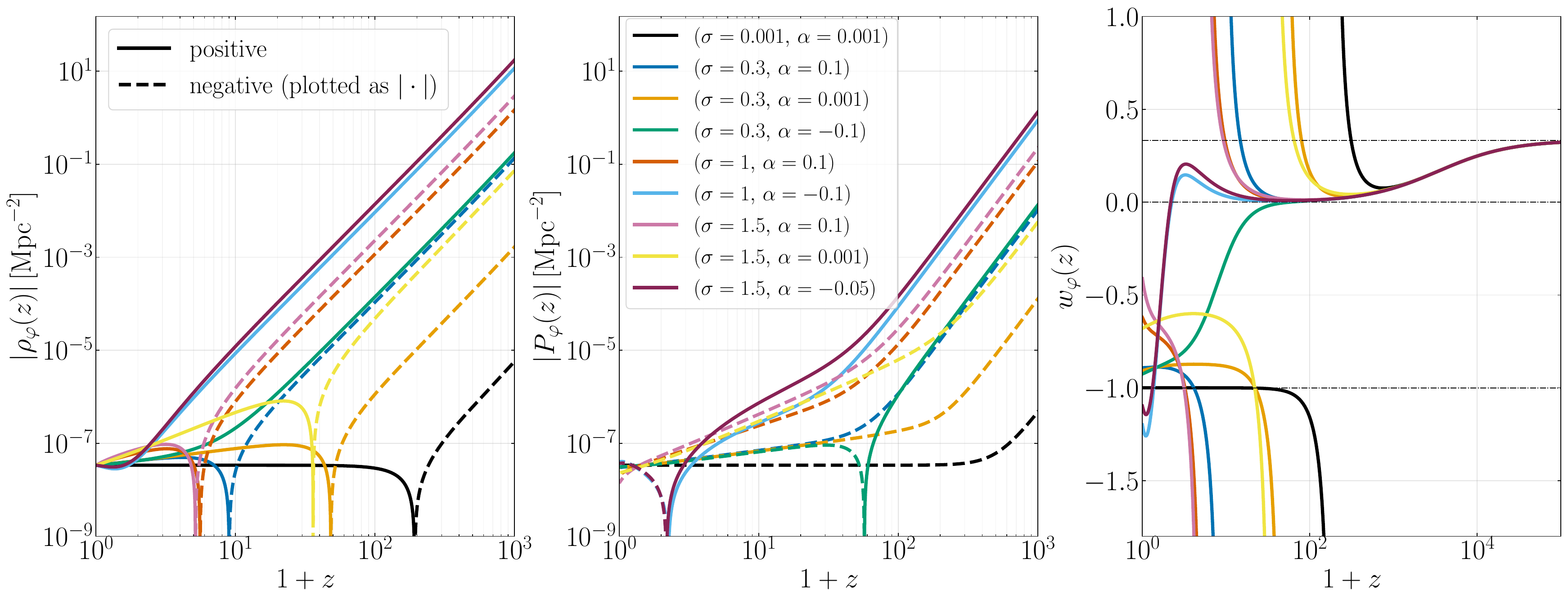}
\caption{
Evolution of the effective scalar-field quantities for several representative parameter choices. 
Left: absolute value of the scalar energy density $|\rho_\varphi(z)|$. 
Middle: absolute value of the scalar pressure $|P_\varphi(z)|$. 
Right: scalar equation of state parameter $w_\varphi(z)$. 
Dashed curves correspond to negative values of the respective quantities. 
The densities follow the standard \texttt{CLASS} normalization (in units of ${\rm Mpc}^{-2}$). 
At early times the scalar contribution is dominated by curvature–coupling terms, which can render $\rho_\varphi$ negative for $\alpha>0$. 
As the scalar field evolves dynamically at late times the potential energy becomes dominant and the equation of state approaches $w_\varphi\approx -1$. 
For positive coupling, the sign change of $\rho_\varphi$ produces a pole
(divergence) in $w_\varphi$, visible in the right panel, rather than a
smooth phantom-divide crossing. By contrast, some negative-coupling cases
exhibit a smooth effective crossing of $w_\varphi=-1$ while
$\rho_\varphi$ remains finite.
}
\label{fig:rho_P_w}
\end{figure}

Further insight into the scalar sector is provided by Fig.~\ref{fig:rho_P_w}, 
which shows the effective scalar-field energy density $\rho_\varphi(z)$, pressure 
$P_\varphi(z)$, and the corresponding equation of state parameter 
$w_\varphi(z)=P_\varphi/\rho_\varphi$. The left and middle panels display the 
absolute values of $\rho_\varphi$ and $P_\varphi$, with dashed curves indicating 
quantities that are negative but plotted in absolute value for clarity. At early 
times the scalar energy density is dominated by the curvature--coupling term, so 
that $\rho_\varphi$ can become negative for $\alpha>0$. 

As the scalar field becomes dynamical at later times, both the kinetic and
potential contributions increase. For the parameter choices considered here,
the potential term nevertheless dominates over the canonical kinetic term, so
the scalar sector enters a potential-dominated regime analogous to slow-roll
quintessence. Consequently, the effective equation of state approaches
$w_\varphi\simeq -1$. The residual departures from this limit are controlled by
the subleading kinetic contribution together with the terms induced by the
non-minimal coupling in Eqs.~\eqref{density}--\eqref{pressure}.

The corresponding evolution of the equation-of-state parameter is shown in the 
right panel of Fig.~\ref{fig:rho_P_w}. In the early Universe the scalar field 
tracks the dominant background component, with
\begin{equation}
w_\varphi \simeq -\frac{\mathcal H^2 + 2{\mathcal H}'}{3\mathcal H^2},
\end{equation}
yielding $w_\varphi\simeq 1/3$ during radiation domination and 
$w_\varphi\simeq 0$ during matter domination. For models with positive coupling 
$\alpha$, the effective scalar density evolves from negative to positive values 
at late times. Since $w_\varphi=P_\varphi/\rho_\varphi$, the zero crossing of 
$\rho_\varphi$ produces a divergence in $w_\varphi$, where it formally evolves 
from $+\infty$ to $-\infty$. The apparent phantom crossing seen in the figure is 
therefore simply a consequence of the 
sign change in the effective scalar energy density, rather than a pathology of the
underlying scalar degree of freedom. 

For negative values of the coupling, however, the behaviour can be qualitatively
different. For example, the models $(\sigma,\alpha)=(1,-0.1)$ and
$(\sigma,\alpha)=(1.5,-0.05)$ exhibit a smooth low-redshift crossing of the
phantom divide. In these cases $w_\varphi$ evolves from the matter-tracking
regime, $w_\varphi\simeq 0$, towards $w_\varphi<-1$ without a divergence, since
the effective scalar density remains finite and does not cross zero in the
relevant redshift interval. This behaviour arises from the non-minimal coupling
terms in the effective fluid description and should therefore be interpreted as
an effective phantom crossing.

\subsubsection{Expansion rate of the Universe}

A direct probe of the background expansion history is provided by measurements
of the expansion rate at different redshifts. In principle, the most direct
determination would come from the so-called redshift drift, or real-time
cosmology, which measures the change of the redshift of a distant comoving
source with respect to the observer's time. In terms of conformal time $\tau$,
the redshift is $1+z=a(\tau_0)/a(\tau_e)$. For two successive photons emitted
and observed along the same pair of comoving worldlines, the comoving distance
to the source is fixed, $\chi_s=\tau_0-\tau_e=\mathrm{constant}$, and therefore
$d\tau_e/d\tau_0=1$. Differentiating the redshift relation with respect to the
observer's conformal time gives
\begin{equation}
\frac{dz}{d\tau_0}
=
(1+z)\left[\mathcal{H}_0-\mathcal{H}(z)\right],
\label{eq:redshift_drift_conformal}
\end{equation}
where $\mathcal{H}\equiv a'/a$, $\mathcal{H}_0\equiv\mathcal{H}(\tau_0)$, and
$\mathcal{H}(z)\equiv\mathcal{H}(\tau_e)$ \citep{Martins:2016bbi}. Although such measurements remain beyond current
observational capabilities, several indirect methods allow the expansion rate
to be estimated with reasonable accuracy.
One of the most robust approaches relies on the so--called \emph{cosmic chronometers}. These are massive, passively evolving galaxies whose differential ages can be inferred from their spectral features. Measuring the age difference of galaxies within a small redshift interval allows a direct determination of $dz/dt$, and therefore of the Hubble parameter via $H(z) = -(1+z)^{-1}dz/dt$ \cite{Jimenez:2001gg}. Independent constraints on the expansion rate can also be obtained from baryon acoustic oscillations (BAO), although these measurements typically assume a specific pre--recombination cosmology.

\begin{figure}[ht]
\centering
\includegraphics[scale=0.55]{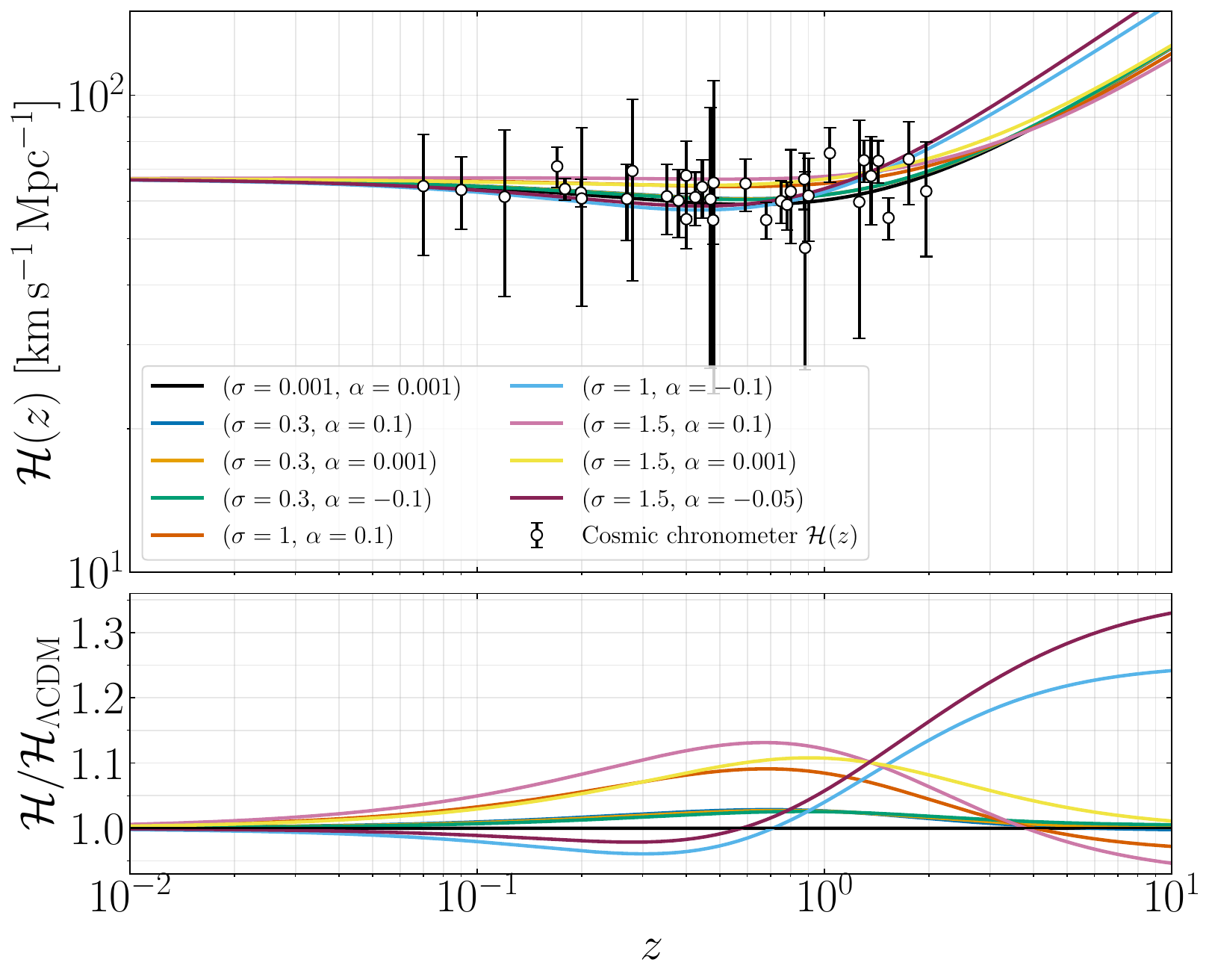}
\caption{
Predicted expansion history for several realizations of the model compared with observational data. 
Top panel: Conformal Hubble parameter $\mathcal H(z)$ in physical units. The points with error bars correspond to cosmic chronometer measurements obtained from 
\cite{Zhang:2012mp,Simon:2004tf,Moresco:2012jh,Moresco:2016mzx,Ratsimbazafy:2017vga,Stern:2009ep,Borghi:2021rft,Jimenez:2023flo,Jiao:2022aep,Tomasetti:2023kek,Moresco:2015cya}, converted to the corresponding conformal Hubble parameter, $\mathcal H=H/(1+z)$.
Bottom panel: ratio of the expansion rate relative to the $\Lambda$CDM prediction, $\mathcal H /\mathcal H_{\Lambda{\rm CDM}}$. 
Different coloured curves correspond to representative choices of the potential slope $\sigma$ and coupling parameter $\alpha$. 
}
\label{fig:hubble}
\end{figure}

To illustrate the effect of the curvature-coupled model on the expansion history, Fig.~\ref{fig:hubble} shows the predicted conformal Hubble parameter for several representative choices of the parameters $(\sigma,\alpha)$ together with cosmic chronometer measurements. The upper panel displays $\mathcal H(z)$ in physical units, while the lower panel shows the ratio $\mathcal H/\mathcal H_{\Lambda{\rm CDM}}$ relative to the $\Lambda$CDM prediction. 

As can be seen, different realizations of the model produce modest but potentially observable deviations from the $\Lambda$CDM expansion history, particularly at intermediate and high redshifts $z\sim 0.5$--$2$. The magnitude and sign of these deviations depend on both the steepness of the scalar potential and the strength of the non--minimal coupling. In general, larger values of $\sigma$ or $|\alpha|$ lead to stronger departures from the standard cosmological expansion rate. Nevertheless, for the representative parameter choices shown here the predicted expansion histories remain broadly consistent with the current cosmic chronometer data.

\subsubsection{Cosmic distances}

Another powerful probe of the expansion history is provided by cosmological 
distance measurements. These rely on objects whose intrinsic properties allow 
their distance to be inferred from observations. Two main classes are commonly 
used: \emph{standard candles}, whose absolute luminosity is known, and 
\emph{standard rulers}, whose physical size is known.

For standard candles, such as Type~Ia supernovae, the observable quantity is the 
luminosity distance $D_{\rm L}(z)$. In a spatially flat universe it is related to 
the expansion history through
\begin{equation}
D_{\rm L}(z) = (1+z)\int_0^z \frac{c\,dz'}{H(z')}.
\end{equation}
The luminosity distance is inferred from the apparent magnitude $m$ of the 
supernovae, which is related to $D_{\rm L}$ through
\begin{equation}
m = M + 5\log_{10}\!\left(\frac{D_{\rm L}}{1\,{\rm Mpc}}\right) + 25 ,
\end{equation}
where $M$ is the calibrated absolute magnitude.

\begin{figure}[ht]
\centering
\includegraphics[scale=0.55]{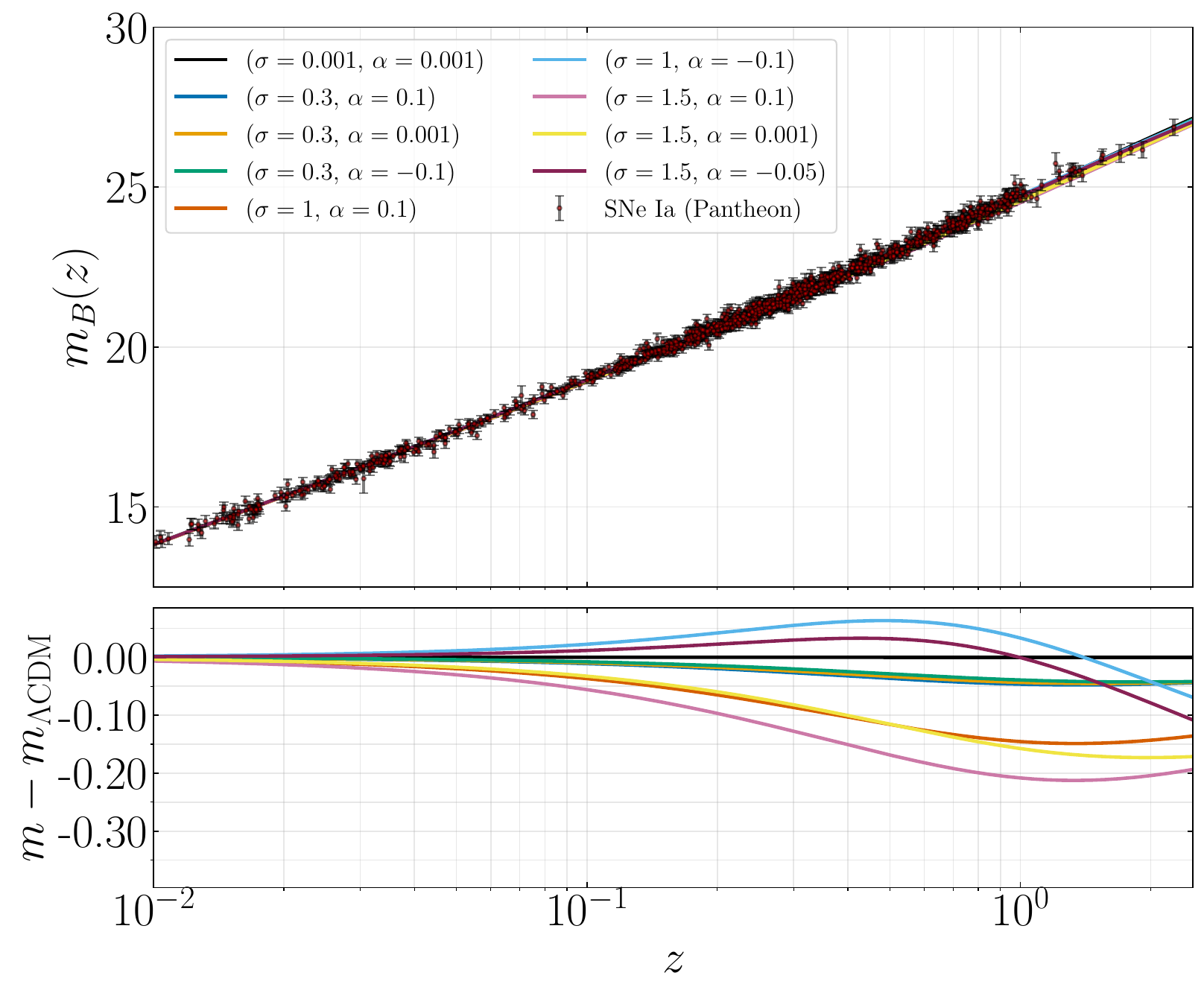}
\caption{
Magnitude--redshift relation for several realizations of the curvature-coupled model compared 
with Type~Ia supernova observations. 
Top panel: apparent magnitude $m_B(z)$ as a function of redshift. The data points 
correspond to the Pantheon supernova compilation \cite{Scolnic:2017caz}. 
Bottom panel: difference relative to the $\Lambda$CDM prediction, 
$m - m_{\Lambda{\rm CDM}}$. 
Different coloured curves correspond to representative choices of the scalar 
potential slope $\sigma$ and the coupling parameter $\alpha$. 
While the models closely follow the $\Lambda$CDM prediction at low redshift, 
larger deviations appear at intermediate redshifts.
}
\label{fig:m_B}
\end{figure}

Figure~\ref{fig:m_B} shows the predicted magnitude--redshift relation for several choices of $\{\alpha,\sigma\}$ in the 
curvature-coupled model together with Type~Ia supernova measurements 
from the Pantheon compilation \cite{Scolnic:2017caz}. The upper panel displays 
the apparent magnitude $m_B(z)$, while the lower panel shows the deviation 
relative to the $\Lambda$CDM prediction. As seen in the figure, the curvature-coupled model 
produces small but potentially observable departures from $\Lambda$CDM, 
particularly at intermediate and high redshifts $z\sim0.5$--$2$. The size of the 
deviation depends on both the steepness of the scalar potential $\sigma$ and 
the strength of the non-minimal coupling $\alpha$. The largest deviations reach $\Delta m \sim -0.20$ at $z\sim1$ for the most extreme parameter choices. For the representative 
parameter choices shown here, however, the predicted distance--redshift relation 
remains broadly consistent with the current supernova data.

\subsection{Linear perturbations}
\label{sec:linear_pert}

We now turn to the linear perturbation sector of the curvature-coupled model. In practice,
the full system is evolved numerically in \texttt{hi\_class}, which solves
the Horndeski perturbation equations after the background evolution and the
corresponding EFT/$\alpha$-functions have been specified from the covariant
Lagrangian in Eqs.~\eqref{eq:G2}-\eqref{eq:eq_lagrangian}.

Unlike the standard implementation of \texttt{hi\_class}, which evolves the
dimensionless variable
$V_X \equiv a\,\delta\varphi/\varphi'$
\citep{Zumalacarregui:2016pph},
we evolve the scalar-field perturbation $\delta\varphi$ directly.
To our knowledge, this is the first use of a $\delta\varphi$-based implementation in \texttt{hi\_class}.
This modification is required because, in the present model,
the background field velocity $\varphi'$ changes sign during the cosmological
evolution (see Fig.~\ref{fig:Omega_i}), making $V_X$ singular even though
$\delta\varphi$ remains regular. This is the same turning-point degeneracy of the unitary-gauge description
discussed in Sec.~\ref{subsec:stability}, and does not correspond to a physical
singularity of the scalar perturbation. For models in which $\varphi'$ does not cross
zero, we verified that the $\delta\varphi$ implementation reproduces the
standard $V_X$ evolution with excellent agreement. The explicit $\delta\varphi$-based perturbation equations are given in
Appendix~\ref{app:pert_eqs}, while the corresponding numerical
implementation in \texttt{hi\_class} is described in
Appendix~\ref{app:hiclass}.

We describe scalar perturbations in the Poisson gauge, for which the line element is,
\begin{equation}
ds^2=a^2(\eta)\left[-(1+2\Psi)d\eta^2+(1-2\Phi)\delta_{ij}dx^i dx^j\right],
\label{eq:pertmetric}
\end{equation}
where $\Phi$ and $\Psi$ are the Bardeen potentials. In the curvature-coupled model, the
non-minimal coupling modifies the linearized Einstein equations as well as
the scalar-field perturbation equation, so that matter perturbations, metric
perturbations and scalar-field fluctuations evolve as a coupled system. At
the physical level, this means that the growth of structure is affected not
only through the modified background expansion, but also through a direct
modification of the relation between matter and the gravitational potentials.

\begin{figure}[ht]
\centering
\includegraphics[scale=0.265]{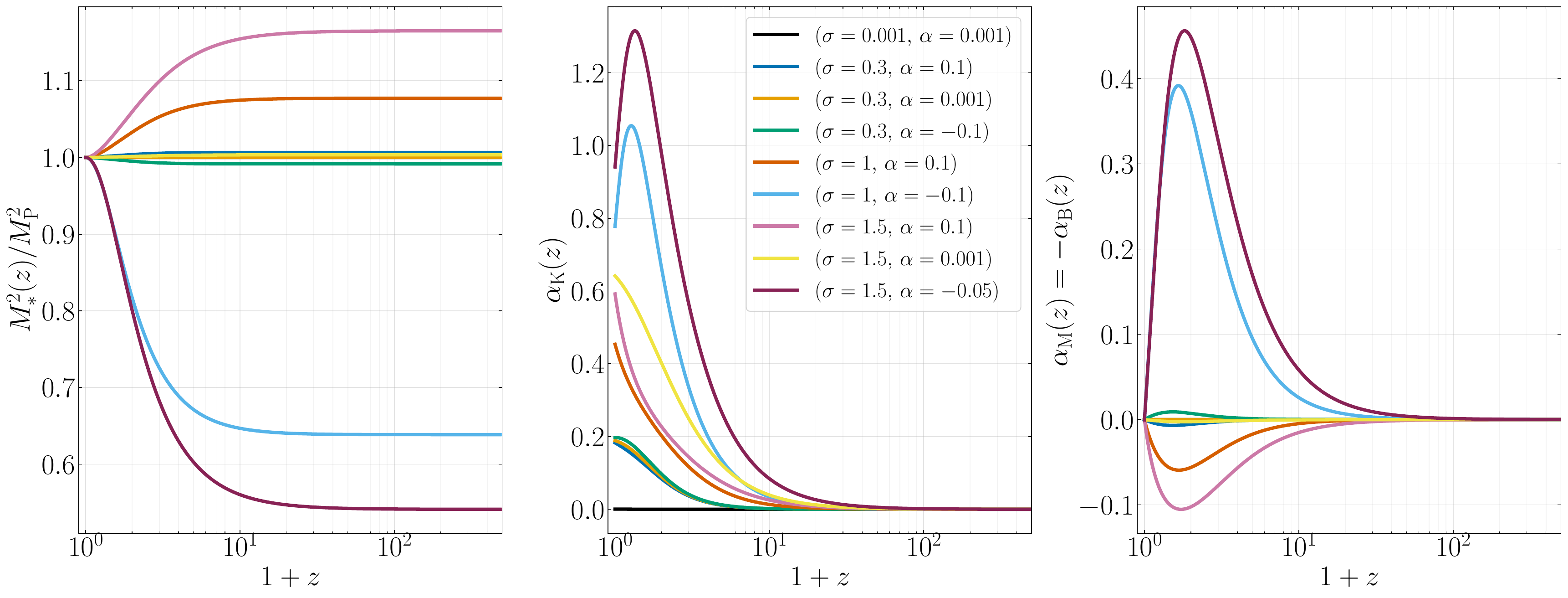}
\caption{
Evolution of the effective Planck mass and EFT/$\alpha$-functions for representative parameters.
The panels show the effective Planck mass $M_*^2$, the kineticity
$\alpha_{\rm K}$, and the Planck-mass run rate $\alpha_{\rm M} = -\alpha_{\rm B}$ as
functions of redshift. The effective Planck mass is plotted in units of $M_{\rm P}^2$,
where $M_*^2/M_{\rm P}^2=1+f(\varphi)$. While the $\alpha_i$ functions shown here are
suppressed at high redshift and become appreciable only at late times, when
the scalar field becomes dynamically relevant, $M_*^2$ can already differ
from its GR value in the early Universe because the shifted non-minimal
coupling remains active throughout the evolution.
}
\label{fig:eft_functions}
\end{figure}

This structure is conveniently described in terms of the effective Planck mass and the EFT/$\alpha$-functions introduced in Eqs.~\eqref{eq:Mstar_2}-\eqref{alpha_i_functions}. Their evolution is shown in Fig.~\ref{fig:eft_functions}. The effective Planck mass $M_*^2$ quantifies the time-dependent normalization of the gravitational sector, while $\alpha_{\rm K}$ and $\alpha_{\rm M}$ describe the kineticity of the scalar perturbation and the running of the Planck mass, respectively. The figure shows two distinct features of the model. First, the $\alpha_i$ functions are strongly suppressed at high redshift and become appreciable only at late times, when the scalar field starts to evolve dynamically. Second, $M_*^2$ need not be close to its GR value at early times, because the shifted coupling keeps $f(\varphi)$ non-zero away from the present epoch, the effective gravitational normalization can be modified even when the time-variation of the scalar field, and hence the $\alpha_i$ functions, remains small. This opens an additional phenomenological channel, since the model can affect the background strength of gravity at early times while producing the main perturbative modifications only at late times. \\
It is worth emphasizing that, in phenomenological EFT parameterizations commonly used in the literature, the $\alpha$-functions are often assumed to scale with the scalar-field fractional density, $\Omega_{\varphi}$, so that perturbative modifications to gravity become relevant only at late times, as expected for a field responsible for late time dark energy. In the present covariant curvature-coupled model, however, this late-time behaviour of the $\alpha_i$ functions is not imposed by hand through a prescribed scaling, but emerges dynamically from the evolution of the scalar field itself, while the effective Planck mass retains an independent evolution controlled by the non-minimal coupling.

\begin{figure}[ht]
\centering
\includegraphics[scale=0.34]{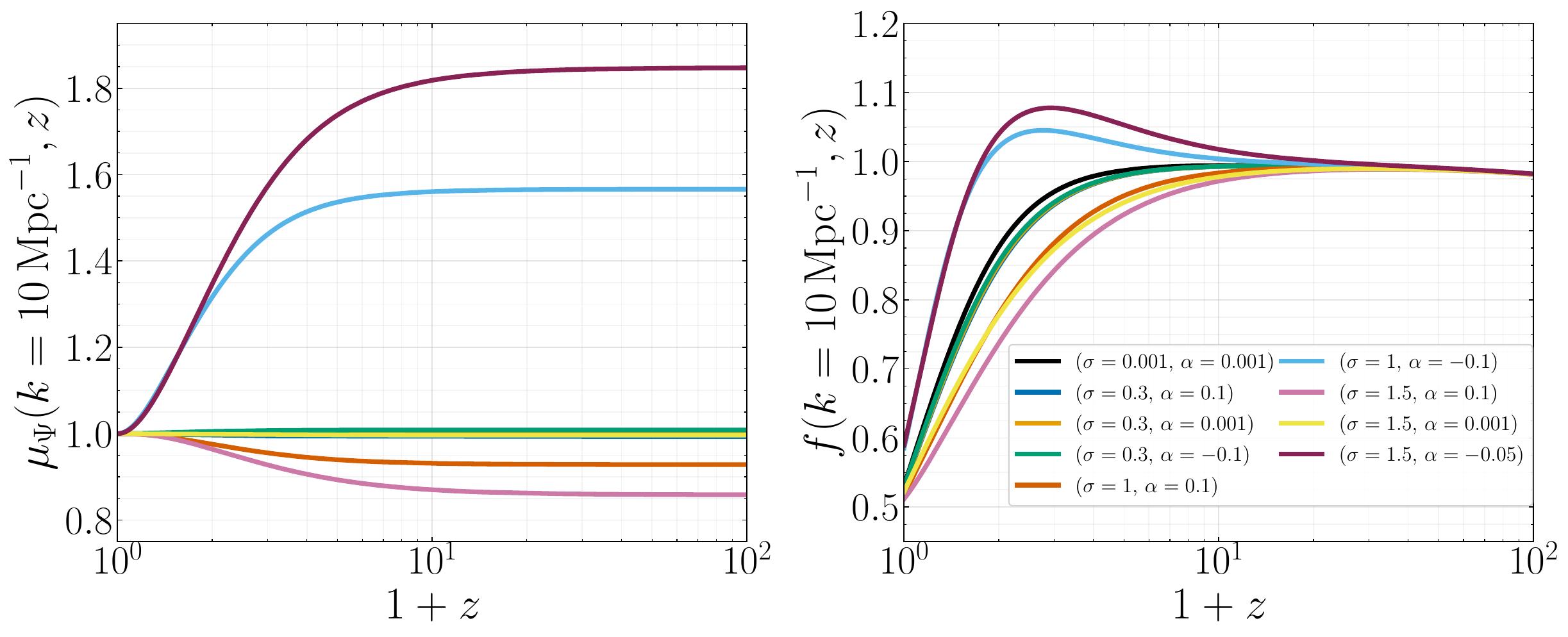}
\caption{
Effective Poisson function and linear growth rate for representative 
parameters. The left panel shows $\mu_\Psi(k,z)$, reconstructed from the full
linear perturbation evolution through Eq.~\eqref{eq:mu_psi_growth}, at
$k=10\,{\rm Mpc}^{-1}$, where it has converged to its quasistatic, high-$k$
limit. The corresponding $G_{\rm eff}/G_N$ obtained from
Eq.~\eqref{eq:Geff} is numerically indistinguishable and is therefore not
shown separately. The right panel shows the linear growth rate,
$f(k,z)=\mathrm d\ln\delta_m/\mathrm d\ln a$, for the same mode.
Negative-$\alpha$, large-$\sigma$ cases enhance the effective gravitational
coupling and structure growth, while positive-$\alpha$ ones generally
suppress both. The case $(\sigma,\alpha)=(0.001,0.001)$ is numerically
indistinguishable from the $\Lambda$CDM prediction over the plotted range.
}
\label{fig:geff_background}
\end{figure}

A complementary characterization of the impact of curvature-coupled model on structure formation is
shown in Fig.~\ref{fig:geff_background}. The \texttt{hi\_class} background
module provides the effective gravitational coupling
$G_{\rm eff}/G_{\rm N}$ using the analytic quasistatic,
$k\rightarrow\infty$ expression of Eq.~\eqref{eq:Geff}. Here, rather than using
this asymptotic quantity directly, we reconstruct the effective Poisson
function from the full numerical evolution of the matter and metric
perturbations by defining
\begin{equation}
-k^2\Psi(k,z)
=
4\pi G_N a^2 \mu_{\Psi}(k,z)\rho_m(z)\delta_m(k,z)\,.
\label{eq:mu_psi_growth}
\end{equation}
Here $\Psi$ is the lapse perturbation in the Poisson gauge and therefore
the potential governing the motion of non-relativistic matter. The
relativistic Poisson equation is sourced by the comoving matter density
perturbation $\Delta_m$. For the scale considered here,
$k=10\,{\rm Mpc}^{-1}$, velocity contributions are negligible, so that
$\Delta_m\simeq\delta_m$ and Eq.~\eqref{eq:mu_psi_growth} accurately
describes the small-scale limit.

The left panel of Fig.~\ref{fig:geff_background} shows
$\mu_\Psi(k,z)$ evaluated at $k=10\,{\rm Mpc}^{-1}$. 
This mode remains well inside the horizon over the entire redshift range shown
and lies sufficiently deep in the high-$k$ regime that $\mu_\Psi$ has reached
its asymptotic quasistatic limit.
As a consistency check, we find
numerical agreement with the effective gravitational coupling computed by
\texttt{hi\_class} from the analytic quasistatic expression,
Eq.~\eqref{eq:Geff}, such that
\begin{equation}
\frac{G_{\rm eff}(z)}{G_N}
\equiv
\lim_{k\rightarrow\infty}\mu_\Psi(k,z).
\label{eq:Geff_highk}
\end{equation}
The curves in the left panel can therefore be interpreted as the effective
gravitational coupling, while being obtained directly from the full linear
perturbation evolution rather than from the analytic quasistatic
approximation. The choice $k=10\,{\rm Mpc}^{-1}$ should thus be understood as
a formal linear-theory diagnostic of the high-$k$ limit, rather than as a
physical scale on which the late-time matter distribution is expected to
remain linear.

The right panel shows the scale-dependent logarithmic growth rate computed
from the same matter perturbation,
\begin{equation}
f(k,z)
\equiv
\frac{\mathrm d\ln\delta_m(k,z)}{\mathrm d\ln a}\,.
\label{eq:growth_rate_delta}
\end{equation}
The two panels provide complementary information. The function $\mu_\Psi$
measures the instantaneous response of the Newtonian potential to the matter
source, whereas $f$ measures the instantaneous rate at which the matter
perturbation evolves. The latter depends not only on the effective
gravitational strength, but also on the expansion history and on the previous
evolution of the coupled matter--metric--scalar system.

All cases recover $\mu_\Psi\simeq1$ at the present epoch, as required by the
shifted coupling and the normalization conditions discussed in
Sec.~\ref{sec:model}. Away from the present epoch, however, different realizations separate according to the evolution of the non-minimal coupling. The
negative-$\alpha$ cases with larger $\sigma$ exhibit the strongest enhancement
of $\mu_\Psi$, corresponding to a stronger effective Newtonian response, while
several positive-$\alpha$ samples give $\mu_\Psi<1$ and hence a weaker response.
This behaviour is consistent with the evolution of the effective Planck mass,
a smaller $M_*^2$ strengthens the metric response, whereas a larger $M_*^2$
suppresses it.

The growth rate follows the same broad ordering, but not in a one-to-one
manner. Models with an enhanced $\mu_\Psi$ generally show faster growth over
part of the evolution, while those with a suppressed effective coupling tend
to grow more slowly. Nevertheless, the differences in $f$ are smaller than
those in $\mu_\Psi$, because the growth rate is also affected by the modified
background expansion. In particular, an enhanced gravitational source can be
partly compensated by stronger Hubble friction. 

\subsubsection{Phenomenological Poisson and slip functions}
\label{subsubsec:poisson_slip}

To characterize the metric response to matter perturbations, we introduce
phenomenological functions modifying the Poisson relations for the scalar
metric potentials. Different conventions are used in the literature,
depending on whether the effective Poisson equation is defined for the spatial
curvature potential $\Phi$ or for the lapse perturbation $\Psi$. A
$\Phi$-based definition is used, for example, in
\cite{Nouri-Zonoz:2024dph}, whereas a $\Psi$-based convention is adopted in
\cite{Planck:2018vyg}. The two definitions coincide in the absence of
gravitational slip, but must be distinguished when $\Phi\neq\Psi$.

In the Poisson gauge convention of Eq.~\eqref{eq:pertmetric}, the time-time
Einstein equation naturally provides a constraint involving $\Phi$, as shown
in Eq.~\eqref{eq:tt_pert_eq_app}. We therefore define
\begin{equation}
-k^2\Phi
=
4\pi G_N a^2\mu_\Phi(k,z)\rho_m\Delta_m .
\label{eq:mu_phi_def}
\end{equation}
Alternatively, the effective Poisson function associated with the lapse
perturbation is defined by
\begin{equation}
-k^2\Psi
=
4\pi G_N a^2\mu_\Psi(k,z)\rho_m\Delta_m .
\label{eq:mu_psi_def}
\end{equation}

The latter definition is directly related to the motion of non-relativistic
matter. For the metric in Eq.~\eqref{eq:pertmetric}, the spatial part of the
geodesic equation, to first order in the metric perturbations and particle
velocity, is
\begin{equation}
\frac{\mathrm d^2x^i}{\mathrm d\tau^2}
+
\mathcal H\frac{\mathrm dx^i}{\mathrm d\tau}
=
-\partial^i\Psi ,
\label{eq:nonrel_geodesic}
\end{equation}
where the force term follows from
$\delta\Gamma^i_{00}=\partial^i\Psi$. Hence, $\Psi$ is the potential governing
the acceleration of non-relativistic tracers, and $\mu_\Psi$ directly
parametrizes the corresponding effective gravitational strength.

The comoving matter density perturbation entering
Eqs.~\eqref{eq:mu_phi_def} and \eqref{eq:mu_psi_def} is
\begin{equation}
\Delta_m
=
\delta_m
+
3\mathcal H\frac{\theta_m}{k^2},
\label{eq:Delta_m_def}
\end{equation}
where $\delta_m$ and $\theta_m$ denote the total density contrast and velocity
divergence of the minimally coupled species entering the Einstein constraint,
excluding the scalar-field perturbation. At late times, this contribution is
effectively dominated by the matter sector, since radiation is negligible.

We define the gravitational slip parameter as
\begin{equation}
\gamma(k,z)
\equiv
\frac{\Phi}{\Psi}.
\label{eq:gamma_def}
\end{equation}
The two Poisson functions then satisfy
\begin{equation}
\mu_\Phi(k,z)
=
\gamma(k,z)\mu_\Psi(k,z).
\label{eq:mu_relation}
\end{equation}
They are therefore not independent once the gravitational slip is specified.

\begin{figure}[ht]
\centering
\includegraphics[scale=0.30]{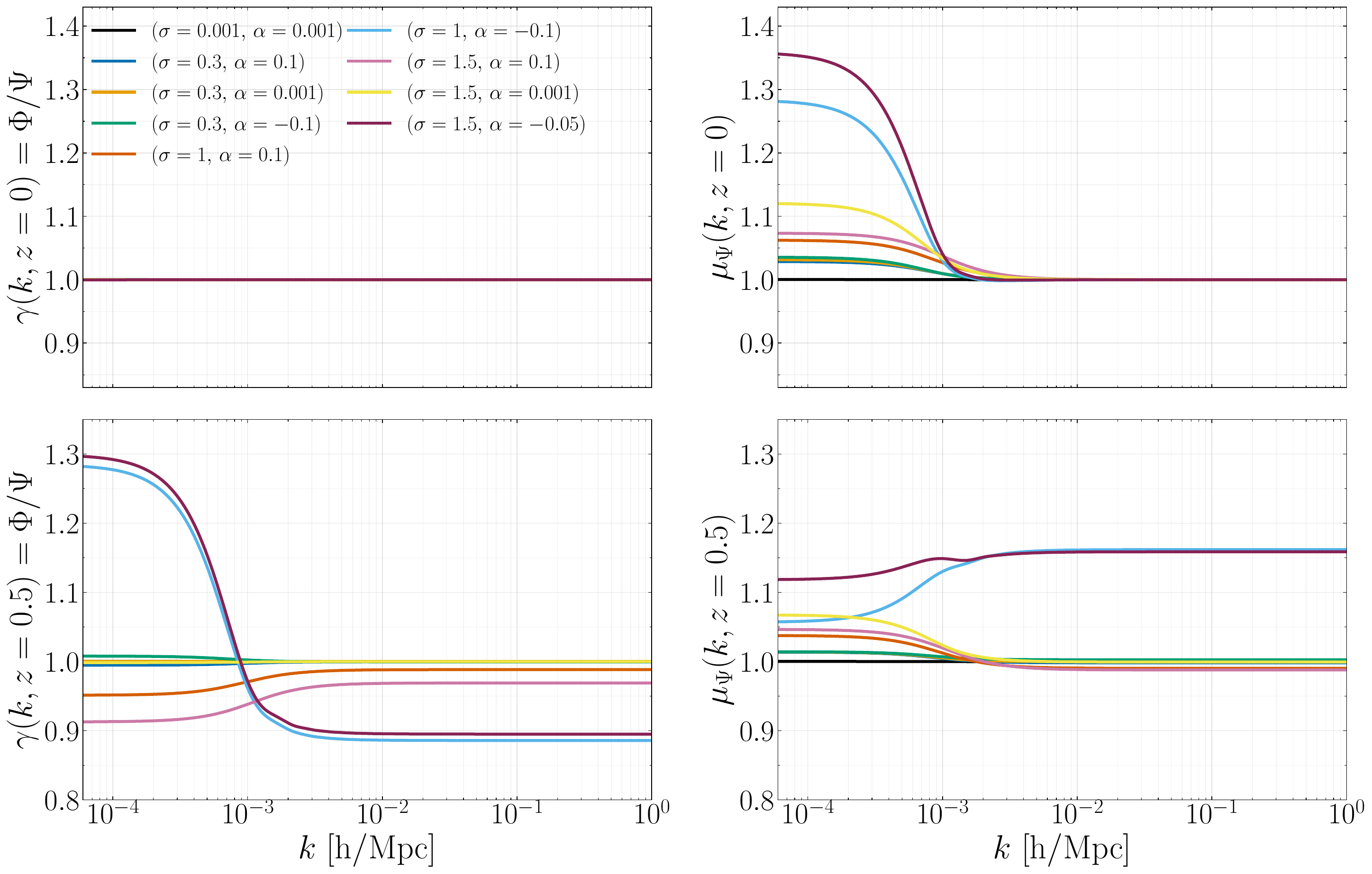}
\caption{
Phenomenological Poisson and slip functions for representative parameters at
$z=0$ and $z=0.5$. The left panels show the gravitational slip
$\gamma(k,z)=\Phi/\Psi$, while the right panels show the effective Poisson
function $\mu_\Psi(k,z)$ defined in Eq.~\eqref{eq:mu_psi_def}. Both functions
approach their GR values in the high-$k$ limit at $z=0$. Toward lower wavenumbers, the scalar field induces model-dependent deviations
in the effective Newtonian response at $z=0$, while the gravitational slip
remains equal to its GR value. At $z=0.5$, deviations appear in both
$\mu_\Psi$ and $\gamma$ over a broader range of scales. The case $(\sigma,\alpha)=(0.001,0.001)$ is numerically
indistinguishable from the $\Lambda$CDM prediction over the plotted range.}
\label{fig:mu_gamma}
\end{figure}

Figure~\ref{fig:mu_gamma} shows $\mu_\Psi(k,z)$ and $\gamma(k,z)$ computed
directly from the \texttt{hi\_class} linear transfer functions. The former
quantifies the scale-dependent modification of the Newtonian response felt by
non-relativistic matter, while the latter measures the ratio between the
two scalar metric potentials. The functions $\mu_\Psi$ and $\gamma$ characterize complementary aspects of
the metric response and their combination fixes the Weyl
potential, $(\Phi+\Psi)/2$, which governs the geodesic motion of massless
particles and is probed by weak gravitational lensing and the ISW effect.

At $z=0$, both functions approach their GR values,
$\mu_\Psi\simeq1$ and $\gamma\simeq1$, in the high-$k$ limit. This is
consistent with the normalization conditions discussed in
Sec.~\ref{local_tests}, which restore the present-day gravitational coupling
and post-Newtonian slip.
Toward lower wavenumbers, the models exhibit scale-dependent departures from
unity. The largest deviations occur for the larger values of $\sigma$ and
negative values of $\alpha$ shown in the Fig.~\ref{fig:mu_gamma}, for which
$\mu_\Psi>1$ corresponds to an enhanced response of the Newtonian potential
to matter perturbations. Deviations of $\gamma$ from unity indicate that
$\Phi$ and $\Psi$ evolve differently and consequently modify the Weyl
potential.

At $z=0.5$, deviations in both functions become visible over a broader range
of scales for several parameter choices. The simultaneous scale and redshift
dependence of $\mu_\Psi$ and $\gamma$ is therefore a characteristic signature
of the scalar contribution to the metric sector in the curvature-coupled model.

\subsubsection{The Cosmic Microwave Background}

The cosmic microwave background (CMB) provides one of the most precise probes 
of cosmology, encoding information about both the early Universe and its 
subsequent evolution \cite{Planck:2015xua, Planck:2018vyg}. Temperature 
anisotropies are commonly characterized by the angular power spectrum
\begin{equation}
D_\ell \equiv \frac{\ell(\ell+1)}{2\pi} C_\ell^{\rm TT},
\end{equation}
which quantifies the variance of temperature fluctuations as a function of 
angular scale.

The observed anisotropies arise from several physical contributions, including 
the Sachs--Wolfe, early and late integrated Sachs--Wolfe and 
Doppler effects. In our analysis, all these contributions are 
consistently included in the computation of the CMB spectra.

\begin{figure}[ht]
\centering
\includegraphics[scale=0.55]{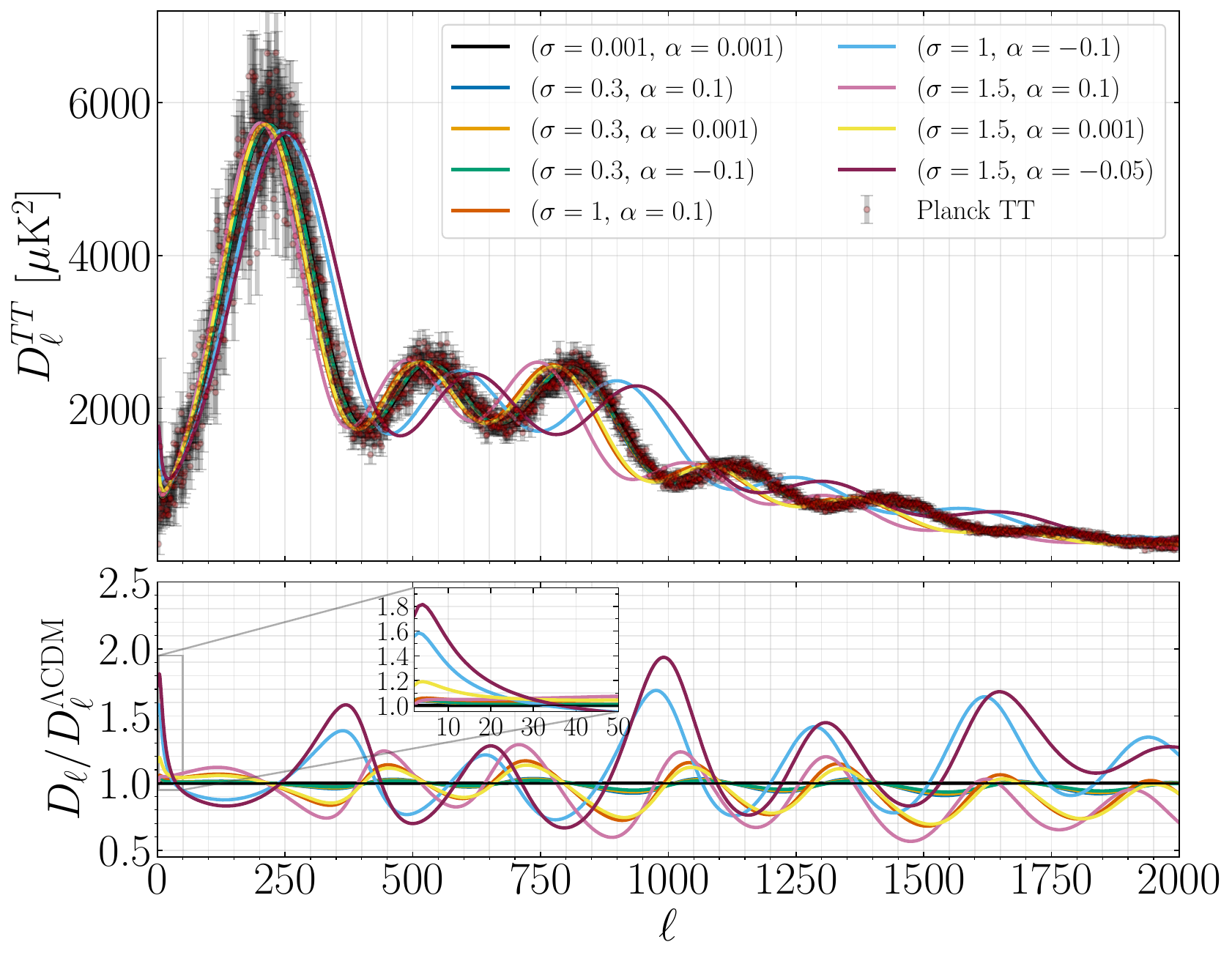}
\caption{
CMB temperature angular power spectrum $D_\ell^{ \rm TT}$ for representative parameters of the curvature-coupled
model compared with unbinned Planck 2018 data \citep{Planck:2019nip}. The lower panel shows the ratio
with respect to the $\Lambda$CDM prediction. The largest deviations occur at low
multipoles, where the spectrum is sensitive to the late-time evolution of the
gravitational potentials through the ISW effect. At higher multipoles, the main
visible effect is a parameter-dependent shift of the acoustic peaks, sourced by the
combined change in the sound horizon at decoupling and the distance to last
scattering surface.
}
\label{fig:D_ell}
\end{figure}

Figure~\ref{fig:D_ell} shows the CMB temperature power spectrum for several parameter choices of
the curvature-coupled model compared to Planck 2018 unbinned data \citep{Planck:2019nip}. The upper panel 
displays $D_\ell^{\rm TT}$, while the lower panel shows the ratio with respect to 
$\Lambda$CDM.

The main deviations appear at low multipoles, where the temperature spectrum is
sensitive to the late time evolution of the gravitational potentials through the
ISW effect. This is the regime in which the scalar field becomes dynamically
relevant and modifies both the background expansion and the metric potentials.
At higher multipoles the spectra remain closer to the $\Lambda$CDM prediction,
although the acoustic peaks are shifted for some parameter choices. This shift
reflects the fact that the curvature-coupled model affects not only the late time dark energy sector, but
also the pre-recombination expansion history through the non-minimal coupling.

The relevant physical scale for the acoustic peaks is the comoving sound
horizon at photon decoupling,
\begin{equation}
r_s(z_*)
=
\int_{z_*}^{\infty}\frac{c_s(z)}{H(z)}\,dz ,
\label{eq:sound_horizon}
\end{equation}
where $z_*$ denotes the redshift of last scattering, $H(z)$ is the physical
Hubble rate, and $c_s$ is the sound speed of the tightly coupled
photon--baryon fluid. The latter is given by
\begin{equation}
c_s(z)
=
\frac{1}{\sqrt{3\left[1+R_b(z)\right]}} ,
\qquad
R_b(z)\equiv \frac{3\rho_b(z)}{4\rho_\gamma(z)} ,
\label{eq:sound_speed}
\end{equation}
with $\rho_b$ and $\rho_\gamma$ the baryon and photon background densities.
The corresponding angular acoustic scale is approximately
\begin{equation}
\ell_A
\simeq
\pi \frac{D_M(z_*)}{r_s(z_*)},
\label{eq:acoustic_scale}
\end{equation}
where
\begin{equation}
D_M(z_*)=\int_0^{z_*}\frac{dz}{H(z)}
\label{eq:comoving_distance_lss}
\end{equation}
is the comoving angular-diameter distance to last scattering in a spatially
flat universe. 
Equivalently, $D_M(z_*)=(1+z_*)D_A(z_*)$, where $D_A$ is the usual physical angular-diameter distance.

In the curvature-coupled model, the shifted non-minimal coupling can affect both quantities entering
Eq.~\eqref{eq:acoustic_scale}. The late-time evolution changes the distance
to last scattering, while the early-time contribution of the coupling modifies
the pre-recombination expansion rate and hence the sound horizon itself. This
second effect follows from the early-time behaviour discussed in
Sec.~\ref{BG_analysis}: when the scalar field is frozen, the effective scalar
density is dominated by the curvature-coupling contribution,
$\rho_\varphi\simeq -3M_{\rm P}^2 f(\varphi)\mathcal{H}^2/a^2$. Since
$f(\varphi_{\rm ini})\simeq \alpha\varphi_{\rm today}^4$ for
$\varphi_{\rm ini}\ll \varphi_{\rm today}$, positive $\alpha$ gives a negative
effective scalar contribution at early times, while negative $\alpha$ gives a
positive one.

As a result, for $\alpha>0$ the effective early-time scalar contribution is
negative, which tends to lower $H(z)$ before recombination and increase the
sound horizon $r_s(z_*)$. For $\alpha<0$, the effective scalar contribution has
the opposite sign, tending to increase $H(z)$ and reduce $r_s(z_*)$. Since the
acoustic scale is controlled by the ratio $D_M(z_*)/r_s(z_*)$, a larger sound
horizon shifts the acoustic peaks toward lower multipoles if the distance to
last scattering is held fixed, while a smaller sound horizon shifts them toward
higher multipoles. In the full calculation of the curvature-coupled model, however, $D_M(z_*)$ also changes
because the late-time expansion history is modified. The peak positions in
Fig.~\ref{fig:D_ell} therefore reflect the combined effect of the early-time
change in $r_s(z_*)$ and the integrated change in $D_M(z_*)$, which is why both
the sign and the size of the shifts depend on $\alpha$ and $\sigma$.

In the present illustrative spectra, $H_0$ is kept fixed. Therefore, any
displacement of the acoustic peaks in Fig.~\ref{fig:D_ell} should be interpreted
as the shift produced by the model parameters at fixed background normalization,
rather than as the result of a parameter fit to the CMB. In a full inference,
$H_0$ would vary together with $\alpha$, $\sigma$, and the physical matter
densities. This matters because the observed acoustic scale tightly constrains
the ratio $D_M(z_*)/r_s(z_*)$, rather than either quantity separately.

The role of $H_0$ should not be interpreted as a pure rescaling of the expansion
rate at all redshifts. Writing
\begin{equation}
H(z)=H_0 E(z),
\end{equation}
a pure rescaling would correspond to varying $H_0$ while keeping the
dimensionless expansion history $E(z)$ fixed. In this artificial limit, both
$D_M(z_*)$ and $r_s(z_*)$ scale approximately as $H_0^{-1}$, so their ratio
remains nearly unchanged. In a cosmological parameter fit, however, changing
$H_0$ is not equivalent to multiplying the entire expansion history by a
constant. The CMB primarily constrains the physical matter densities
$\omega_b\equiv\Omega_b h^2$ and $\omega_c\equiv\Omega_c h^2$, together with
the angular acoustic scale. The measured CMB monopole fixes the present photon
temperature and hence $\omega_\gamma$, while the total radiation density
$\omega_r$ is then determined once the neutrino sector, in particular
$N_{\rm eff}$, is specified.  At high
redshift, the terms entering the expansion rate are of the form
$H_0^2\Omega_i$, which can be written in terms of the physical densities
$\omega_i$. Hence, once $\omega_b$ and $\omega_c$ are fixed, changing $h$ has
only a limited direct effect on the pre-recombination expansion rate and
therefore on $r_s(z_*)$. Instead, changing $H_0$ mainly changes the fractional
densities $\Omega_i=\omega_i/h^2$ and the late-time matter--dark-energy balance,
thereby modifying the distance $D_M(z_*)$. The sound horizon is therefore most
sensitive to the physical pre-recombination energy budget and to genuine
early-time modifications of $H(z)$, such as the non-minimal coupling contribution in the curvature-coupled model, while
$H_0$ primarily acts through the distance to last scattering.

Consequently, in a parameter fit, changes in $H_0$ mainly compensate changes in
$r_s(z_*)$ by adjusting the distance $D_M(z_*)$. This is particularly relevant
for the negative-$\alpha$ models, for which the early-time effective scalar
contribution tends to increase the pre-recombination expansion rate and reduce
the sound horizon. At fixed $H_0$, this moves the acoustic scale in the same
qualitative direction as early-time mechanisms designed to alleviate the Hubble
tension. A smaller $r_s(z_*)$ can be reconciled with the observed peak positions
by a larger inferred $H_0$, through the corresponding reduction of
$D_M(z_*)$. Determining whether this compensation is viable in practice requires a full
likelihood analysis, since the same parameter changes that restore the acoustic
scale also affect the peak heights, damping tail, CMB lensing amplitude, and
late-time observables.

\subsubsection{Power spectra}

The curvature-coupled model modifies both the background expansion and the linear evolution of
matter and metric perturbations, leaving scale-dependent signatures in the power
spectra. We compute these spectra directly from the \texttt{hi\_class} transfer
functions. Following the definition used in \cite{Lesgourgues:2011re}, for any scalar perturbation variable $X$, the transfer function $T_X(k,z)$
relates the primordial curvature perturbation to the late-time perturbation,
\begin{equation}
X(\mathbf{k},z)
=
T_X(k,z)\,\mathcal R(\mathbf{k}) .
\end{equation}
Here $\mathcal R(\mathbf{k})$ is the primordial comoving curvature perturbation,
and $k=|\mathbf{k}|$. Its dimensionless power spectrum is defined by
\begin{equation}
\left\langle \mathcal R(\mathbf{k})\mathcal R^*(\mathbf{k}')\right\rangle
=
(2\pi)^3\delta_D(\mathbf{k}-\mathbf{k}')
\frac{2\pi^2}{k^3}\mathcal{P}_{\mathcal R}(k),
\label{eq:primordial_spectrum_def}
\end{equation}
and is parametrized as
\begin{equation}
\mathcal{P}_{\mathcal R}(k)
=
A_s\left(\frac{k}{k_p}\right)^{n_s-1},
\label{eq:primordial_spectrum_param}
\end{equation}
where $A_s$ is the scalar amplitude, $n_s$ is the scalar spectral index, and
$k_p$ is the pivot scale.

The dimensionless power spectrum of $X$ is therefore
\begin{equation}
\mathcal{P}_X(k,z)
=
\mathcal{P}_{\mathcal R}(k)\,|T_X(k,z)|^2 .
\end{equation}
Equivalently, $\mathcal{P}_X(k,z)$ gives the contribution of each logarithmic
interval in $k$ to the variance of $X$. For the matter sector we show the usual
dimensionful matter power spectrum,
\begin{equation}
P_{\rm m}(k,z)
=
\frac{2\pi^2}{k^3}\mathcal{P}_{\rm m}(k,z),
\end{equation}
while for the metric sector we present the dimensionless Weyl-potential
spectrum. It is worth emphasizing that all results presented here are obtained within
linear perturbation theory, so the results extending into non-linear scales should
 be interpreted as formal linear-theory continuations.

\begin{figure}[ht]
\centering
\includegraphics[scale=0.375]{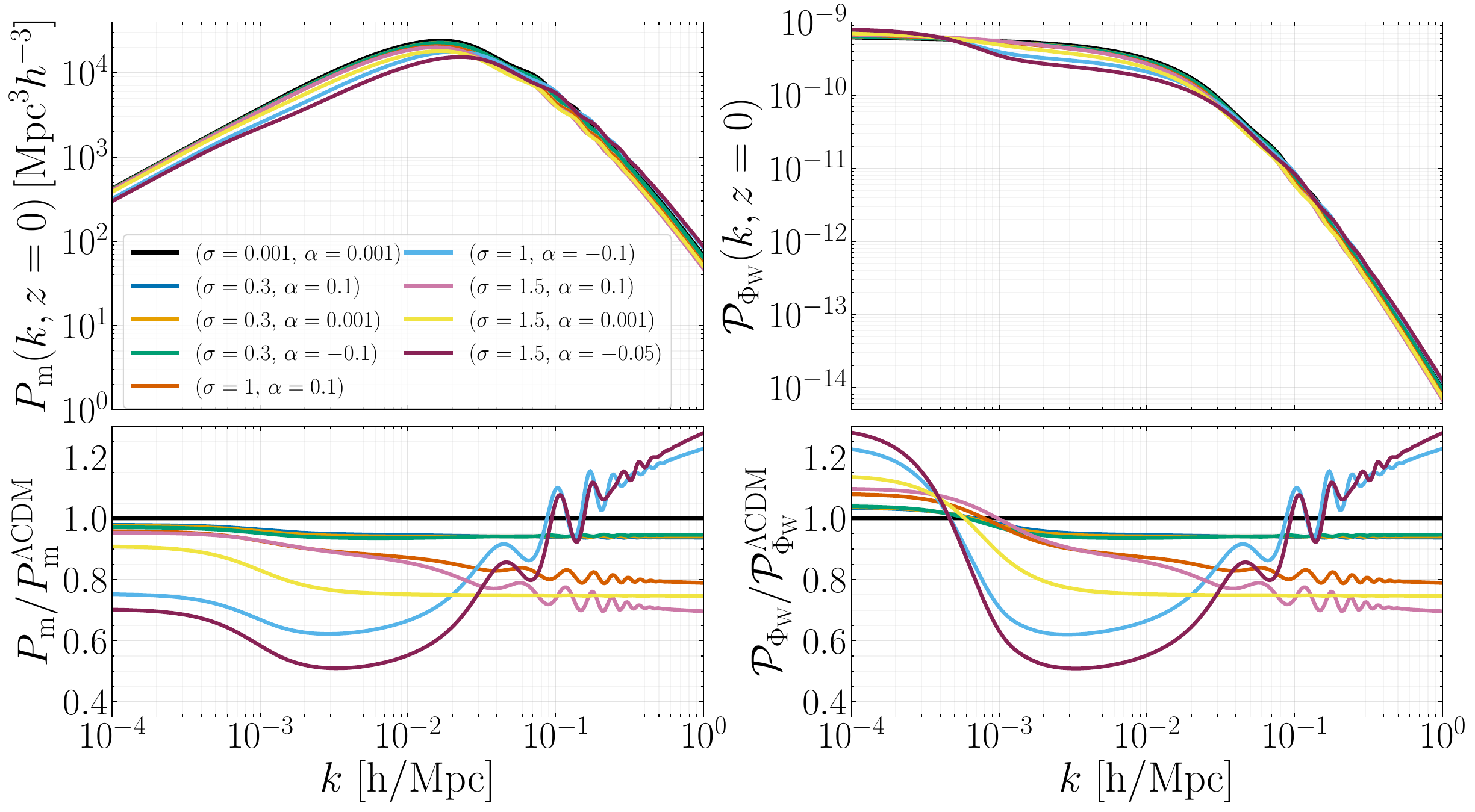}
\caption{
Linear matter power spectrum $P_{\rm m}(k)$ and dimensionless Weyl-potential
spectrum $\mathcal{P}_{\Phi_W}(k)$ at $z=0$ for representative model parameters, with
$\Phi_W=(\Phi+\Psi)/2$. The lower panels show the ratio with respect to
the $\Lambda$CDM prediction. On
quasi-linear and smaller scales the matter and Weyl ratios follow similar
trends, while on the largest scales the Weyl spectrum differs more visibly
because it depends directly on the metric potentials and their evolution.
}
\label{fig:pk_mm_phi}
\end{figure}

Figure~\ref{fig:pk_mm_phi} shows the matter and Weyl-potential spectra at
$z=0$. The left panel displays the linear matter power spectrum $P_{\rm m}(k)$,
which quantifies the clustering amplitude of matter density perturbations at
the present epoch. The right panel shows the dimensionless Weyl-potential
spectrum $\mathcal{P}_{\Phi_W}(k)$, with
$\Phi_W=(\Phi+\Psi)/2$. This quantity probes the metric sector directly and is
therefore relevant for observables such as weak lensing and the ISW effect.

The curvature-coupled model produces scale-dependent deviations in both spectra. Over the
quasi-linear and smaller scales shown in Fig.~\ref{fig:pk_mm_phi}, the ratios
of the matter and Weyl-potential spectra follow broadly similar trends, whereas
their behaviour differs more clearly on the largest scales. The similarity on
sub-horizon scales follows from the fact that the metric potentials are then
approximately determined by the matter perturbation through the modified
Poisson and slip relations. In this regime the velocity contribution to the
comoving density perturbation is suppressed, so that
$\Delta_m\simeq\delta_m$, and the modified Poisson equation,
Eq.~\eqref{eq:mu_psi_def}, together with the slip definition,
Eq.~\eqref{eq:gamma_def}, gives
\begin{equation}
-k^2\Phi_W
=
4\pi G_N a^2
\Sigma(k,z)\rho_m\delta_m,
\qquad
\Sigma(k,z)\equiv
\frac{1+\gamma(k,z)}{2}\,\mu_\Psi(k,z).
\label{eq:weyl_sigma_relation}
\end{equation}
Thus, on sub-horizon scales the Weyl potential is
tied to the same matter perturbation that sources the matter power spectrum, up
to the response factor $\Sigma(k,z)$ and the usual Poisson factor $k^{-2}$.
This relation also explains the similarity between the two ratios at large
$k$. At $z=0$, Fig.~\ref{fig:mu_gamma} shows that $\mu_\Psi\simeq 1$ and
$\gamma\simeq 1$ on sufficiently small scales, as expected from the present-day
local normalization of the model. Hence $\Sigma(k,z=0)\simeq1$, and the
instantaneous mapping between $\Phi_W$ and $\delta_m$ becomes GR-like. This does
not imply that $\mathcal{P}_{\Phi_W}$ itself is identical to the
$\Lambda$CDM prediction, because $\delta_m$ has evolved differently in the curvature-coupled model.
Rather, it implies that the ratio
$\mathcal{P}_{\Phi_W}/\mathcal{P}_{\Phi_W}^{\Lambda {\rm CDM}}$ mainly inherits the
scale dependence of $P_m/P_m^{\Lambda {\rm CDM}}$ whenever $\Sigma/\Sigma^{\Lambda {\rm CDM}}$ is
close to unity. On horizon-scale modes this quasistatic argument no longer
applies: time derivatives of the metric potentials, scalar-field perturbations,
gravitational slip, and the modified background evolution all enter the full
linear solution, leading to the larger difference between the matter and Weyl
spectra on the largest scales.

We now use the matter spectrum to isolate the part of the response most directly
associated with the growth of density perturbations. Although the spectra shown
in Fig.~\ref{fig:pk_mm_phi} are obtained from the full \texttt{hi\_class}
evolution, the sub-horizon limit provides a useful analytic guide for
interpreting the smaller-scale behaviour. Since matter is minimally coupled in
the Jordan frame, its stress-energy tensor is covariantly conserved. For
pressureless matter in the Poisson gauge metric of Eq.~\eqref{eq:pertmetric},
the linear continuity and Euler equations are \citep{Ma:1995ey}
\begin{align}
&\delta_m'
=
-\theta_m
+
3\Phi',
\label{eq:matter_continuity}
\\
& \theta_m'
+
\mathcal H\theta_m
=
k^2\Psi ,
\label{eq:matter_euler}
\end{align}
where $\theta_m\equiv \partial_i v_m^i$ is the divergence of the matter peculiar
velocity, equivalently $\theta_m=i k_i v_m^i$ in Fourier space, and a prime
denotes a derivative with respect to conformal time. The continuity equation
describes the change in the matter density contrast due to velocity convergence
and the perturbation of the spatial volume, while the Euler equation shows that
non-relativistic matter is accelerated by the lapse perturbation $\Psi$.

Combining Eqs.~\eqref{eq:matter_continuity} and \eqref{eq:matter_euler} gives
\begin{equation}
\delta_m''
+
\mathcal H\delta_m'
+
k^2\Psi
=
3\left(\Phi''+\mathcal H\Phi'\right).
\label{eq:delta_exact_intermediate}
\end{equation}
In the sub-horizon, quasistatic regime, the time derivatives of the metric
potentials are subdominant compared with the spatial-gradient term. Using the
$\Psi$-based modified Poisson equation, Eq.~\eqref{eq:mu_psi_def}, and taking
$\Delta_m\simeq\delta_m$ in this limit, Eq.~\eqref{eq:delta_exact_intermediate}
reduces to
\begin{equation}
\delta_m''
+
\mathcal H\delta_m'
-
4\pi G_N a^2 \mu_\Psi(k,a)\rho_m\delta_m
\simeq
0 .
\label{eq:growth_qs_conformal}
\end{equation}
This equation is not used in the numerical calculation, where the full coupled
linear system is evolved with \texttt{hi\_class}. Nevertheless, it provides a
useful interpretation of the smaller-scale trends: matter growth is controlled
by the competition between the friction term set by the expansion history and
the effective gravitational source term proportional to $\mu_\Psi\rho_m$.

The large-scale suppression visible in the matter spectrum should be interpreted
with more care than the smaller-scale trends. These modes are closer to the
horizon scale, where the quasistatic approximation in
Eq.~\eqref{eq:growth_qs_conformal} is not reliable and the density perturbation
retains sensitivity to the time evolution of the metric potentials, the scalar
field perturbation, and the modified background history. Moreover,
$P_m(k,z=0)$ reflects the full transfer-function evolution from the primordial
initial conditions to the present epoch, rather than the instantaneous value of
$\mu_\Psi(k,z=0)$. The suppression on large and intermediate scales should
therefore be viewed as the net response of the full linear system.

For the positive-coupling models ($\alpha >0$), the suppression of $P_m(k)$ is consistent
with this sub-horizon intuition. Away from the present epoch,
$f(\varphi)>0$ increases the effective Planck mass,
$M_*^2=M_{\rm P}^2[1+f(\varphi)]$, and therefore tends to weaken the metric response
to matter sources. This is also reflected in the weak-field expression for the
effective gravitational coupling, Eq.~\eqref{eq:Geff}, where the leading factor
scales as $1/[1+f(\varphi)]$. In addition, these cases exhibit an enhanced
expansion rate relative to the $\Lambda$CDM model over part of
the redshift range relevant for structure formation (Fig.~\ref{fig:hubble}), which increases the
friction term in Eq.~\eqref{eq:growth_qs_conformal}. Both effects act in the
direction of reducing the growth of matter perturbations.

For the negative-coupling cases ($\alpha <0$) with larger $\sigma$, the matter spectrum is
more scale dependent. The power is suppressed on large and intermediate scales,
but becomes enhanced toward the smaller scales shown in the figure. This is
consistent with the enhanced effective coupling in the $\Psi$-Poisson relation
visible in Fig.~\ref{fig:mu_gamma}. The correspondence is not one-to-one, as
$P_m(k,z=0)$ is not determined only by the instantaneous value of
$\mu_\Psi(k,z=0)$, but by the full prior evolution of the coupled
matter--metric--scalar system. The large-scale suppression should therefore be
interpreted as the integrated response of the full linear dynamics, where
metric time derivatives, gravitational slip, scalar-field perturbations, and
the modified background evolution can all be relevant. This also explains why
the Weyl spectrum departs more clearly from the matter spectrum on the largest
scales: unlike $P_m$, it probes the metric potentials directly and is therefore
more sensitive to relativistic metric evolution and gravitational slip.

\subsubsection{Weak gravitational lensing}
\label{subsubsec:lensing}

Weak gravitational lensing probes the metric sector through the deflection
and distortion of photon bundles by the inhomogeneous gravitational field
along the line of sight.
In the Poisson gauge, scalar perturbations are described by the two
Bardeen potentials $\Phi$ and $\Psi$, as in
Eq.~\eqref{eq:pertmetric}. At linear order, and within the Born
approximation, photon deflections are sourced by the Weyl potential,
\begin{equation}
\Phi_{\rm W}
\equiv
\frac{\Phi+\Psi}{2}.
\end{equation}
For a source plane at comoving distance $\chi_s$, the deflection angle is
\cite{Lewis:2006fu}
\begin{equation}
\boldsymbol{\alpha}\bigl(\hat{\mathbf n},z_s\bigr)
=
-2\int_0^{\chi_s}\mathrm d\chi\,
\frac{\chi_s-\chi}{\chi_s\,\chi}\,
\nabla_{\!\hat{\mathbf n}}
\Phi_{\rm W}
\bigl(\chi\hat{\mathbf n},\eta_0-\chi\bigr),
\label{eq:deflection_dollars2}
\end{equation}
where $\hat{\mathbf n}$ is the observed line-of-sight direction and
$\nabla_{\!\hat{\mathbf n}}$ is the angular gradient on the unit sphere.
The metric potentials are evaluated along the unperturbed background
light ray, $\mathbf x(\chi)=\chi\hat{\mathbf n}$ and
$\eta(\chi)=\eta_0-\chi$. The overall sign follows from our convention for the angular displacement
$\boldsymbol{\alpha}$ and from the metric potential convention in
Eq.~\eqref{eq:pertmetric}; the same convention is used consistently in the
definitions of $\phi^{\rm L}$, $\kappa$, and $\mathcal E_{ab}$ below.  Further details on the geometrical origin of
the kernel and the endpoint behaviour are given
in Appendix~\ref{app:lensing_geometry}.

It is useful to introduce the scalar lensing potential $\phi^{\rm L}$ through 
\begin{equation}
\boldsymbol{\alpha}\bigl(\hat{\mathbf n},z_s\bigr)
=
\nabla_{\!\hat{\mathbf n}}
\phi^{\rm L}(\hat{\mathbf n},z_s).
\end{equation}
Comparing with Eq.~\eqref{eq:deflection_dollars2} results in
\begin{equation}
\phi^{\rm L}(\hat{\mathbf n},z_s)
=
-2\int_0^{\chi_s}\mathrm d\chi\,
\frac{\chi_s-\chi}{\chi_s\,\chi}\,
\Phi_{\rm W}
\bigl(\chi\hat{\mathbf n},\eta_0-\chi\bigr).
\label{eq:lenspot_full_dollars2}
\end{equation}
The observable weak-lensing information is contained in the angular
variation of the deflection. This is described by the distortion tensor, its trace or namely convergence, and its trace free part which corresponds to shear. These are defined and derived in Appendix~\ref{app:lensing_distortion}.
We now move from the real-space lensing fields to their angular spectra.
Since $\phi^{\rm L}$ is a scalar on the sky, we expand it in ordinary spherical
harmonics,
\begin{equation}
\phi^{\rm L}(\hat{\mathbf n},z_s)
=
\sum_{\ell m}
\phi^{\rm L}_{\ell m}(z_s)Y_{\ell m}(\hat{\mathbf n}),
\end{equation}
Using the relation between the convergence and the lensing potential in Eq.~\eqref{eq:kappa_lenspot_dollars2}, together with
$\nabla_{\!\hat{\mathbf n}}^2Y_{\ell m}
=-\ell(\ell+1)Y_{\ell m}$, we obtain
\begin{equation}
\kappa_{\ell m}(z_s)
=
\frac{1}{2}
\ell(\ell+1)
\phi^{\rm L}_{\ell m}(z_s).
\end{equation}
For any projected scalar field $X(\hat{\mathbf n},z_s)$, statistical
isotropy implies that its two-point function in harmonic space is diagonal,
allowing the angular power spectrum to be defined as
\begin{equation}
\left\langle
X_{\ell m}(z_s)X^\ast_{\ell' m'}(z_s)
\right\rangle
=
\delta_{\ell\ell'}\delta_{mm'}\,
C_\ell^{XX}(z_s).
\end{equation}
Therefore
\begin{equation}
C_\ell^{\kappa\kappa}(z_s)
=
\frac{1}{4}
\ell^2(\ell+1)^2
C_\ell^{\phi^{\rm L}\phi^{\rm L}}(z_s).
\end{equation}
At leading order in scalar perturbations, the shear contains only an
$E$-mode and no $B$-mode. Since the shear $E$-mode is fixed by the same
lensing potential as the convergence, it carries no independent information
at this order \citep{2013PhDT.......125B, Hassani:2020buk}; we therefore focus on the convergence spectrum. 

The full-sky convergence spectrum can be written directly in terms of the
Weyl-potential power spectrum as
\begin{align}
C_{\ell}^{\kappa\kappa}(z_s)
=
\frac{2}{\pi}\,\ell^2(\ell+1)^2
&\int_0^\infty \mathrm d k\,k^2
\int_0^{\chi_s}
\frac{\mathrm d\chi}{\chi}\,
\frac{\chi_s-\chi}{\chi_s}
\int_0^{\chi_s}
\frac{\mathrm d\chi'}{\chi'}\,
\frac{\chi_s-\chi'}{\chi_s}
\nonumber\\
&\times
j_\ell(k\chi)\,
j_\ell(k\chi')\,
P_{\Phi_{\rm W}}(k;\chi,\chi') .
\label{eq:Cl_kappa_full_Phi_dollars2}
\end{align}
Here $P_{\Phi_{\rm W}}(k;\chi,\chi')$ denotes the unequal-time
Weyl-potential power spectrum, with
$\Phi_{\rm W}(\mathbf k,\chi)\equiv
\Phi_{\rm W}(\mathbf k,\eta_0-\chi)$. It is defined by
\begin{equation}
\left\langle
\Phi_{\rm W}(\mathbf k,\chi)
\Phi_{\rm W}^{\ast}(\mathbf k',\chi')
\right\rangle
=
(2\pi)^3
\delta_{\rm D}(\mathbf k-\mathbf k')\,
P_{\Phi_{\rm W}}(k;\chi,\chi') .
\end{equation}
Equation~\eqref{eq:Cl_kappa_full_Phi_dollars2} keeps the full radial
projection and therefore does not rely on the Limber approximation. The derivation of this expression, the meaning of the unequal-time spectrum,
and the Limber limit are given in Appendix~\ref{app:lensing_limber}.

Figure~\ref{fig:lensing_cls} shows the convergence angular power spectrum
for source planes at $z_s=0.85$ and $z_s=3.3$. These source redshifts are
chosen following \citep{Hassani:2020buk} and allow us to compare lensing by
structures over different portions of the late-time line of sight. Since
weak lensing is sourced by the Weyl potential, the behaviour of
$C_\ell^{\kappa\kappa}$ reflects the scale- and redshift-dependent Weyl
power after projection along the photon path.

At low multipoles, the convergence spectra are generally suppressed
relative to the corresponding $\Lambda$CDM prediction. This should not be
read directly from the low-$k$ behaviour of the equal-time Weyl-potential
spectrum at $z=0$ in Fig.~\ref{fig:pk_mm_phi}. The convergence spectrum is a
line-of-sight projection of Weyl-potential correlations, as shown in
Eq.~\eqref{eq:Cl_kappa_full_Phi_dollars2}, and the lensing efficiency
vanishes at both the observer and the source plane. Therefore the signal is
not dominated by the Weyl potential at either endpoint, but by the integrated
Weyl power over the interior of the photon path. The low-$\ell$ suppression
should therefore be interpreted as a projected effect over a range of
redshifts and scales, rather than as a direct measurement of the present-day
large-scale Weyl spectrum.
This is supported by Fig.~\ref{fig:Weyl_spectra}. At the
redshifts relevant for the lensing projection, most parameter choices show a
suppression of the Weyl-potential power over a broad range of large and
intermediate scales. This suppression feeds into the projected convergence
spectrum and explains why $C_\ell^{\kappa\kappa}$ is reduced at low
multipoles. At these multipoles the Limber approximation is not fully
reliable, so the correspondence $k\simeq(\ell+1/2)/\chi$ should be used only
as a guide: the spherical-Bessel kernels are broad and the signal receives
weighted contributions from a range of wavenumbers and redshifts.

At larger multipoles, the behaviour becomes more model dependent. The
positive-$\alpha$ models remain suppressed over most of the multipole range,
consistent with the suppression of the Weyl-potential power seen in
Fig.~\ref{fig:Weyl_spectra}. By contrast, the negative-$\alpha$ models with
larger $\sigma$ show a scale-dependent transition: they are suppressed at
low $\ell$, but rise above $\Lambda$CDM at high $\ell$. This behaviour
follows from the transition visible in the Weyl-potential spectra. For these
models, $\mathcal P_{\Phi_{\rm W}}/\mathcal P_{\Phi_{\rm W}}^{\Lambda {\rm CDM}}$ is
below unity on large and intermediate scales, but crosses above unity at
smaller scales, around
\begin{equation}
k_{\rm tr}
\sim
{\rm few}\times 10^{-2}\,h\,{\rm Mpc}^{-1},
\end{equation}
with a mild redshift dependence. In the Limber approximation, a feature at
wavenumber $k_{\rm tr}$ contributes most strongly to multipoles
\begin{equation}
\ell_{\rm tr}
\simeq
k_{\rm tr}\,\chi_{\rm eff}(z_s),
\end{equation}
where $\chi_{\rm eff}$ denotes the range of comoving distances to which the
lensing kernel gives significant weight. Since the higher-redshift source
plane samples a longer line of sight and has a larger effective lensing
distance, the same transition in $k$ is shifted to larger multipoles for
$z_s=3.3$ than for $z_s=0.85$. This is precisely the trend seen in
Fig.~\ref{fig:lensing_cls}: the negative-$\alpha$ models cross from
suppression to enhancement at lower $\ell$ for the lower-redshift source
plane, and at higher $\ell$ for the higher-redshift source plane.

\begin{figure}[ht]
\centering
\includegraphics[scale=0.26]{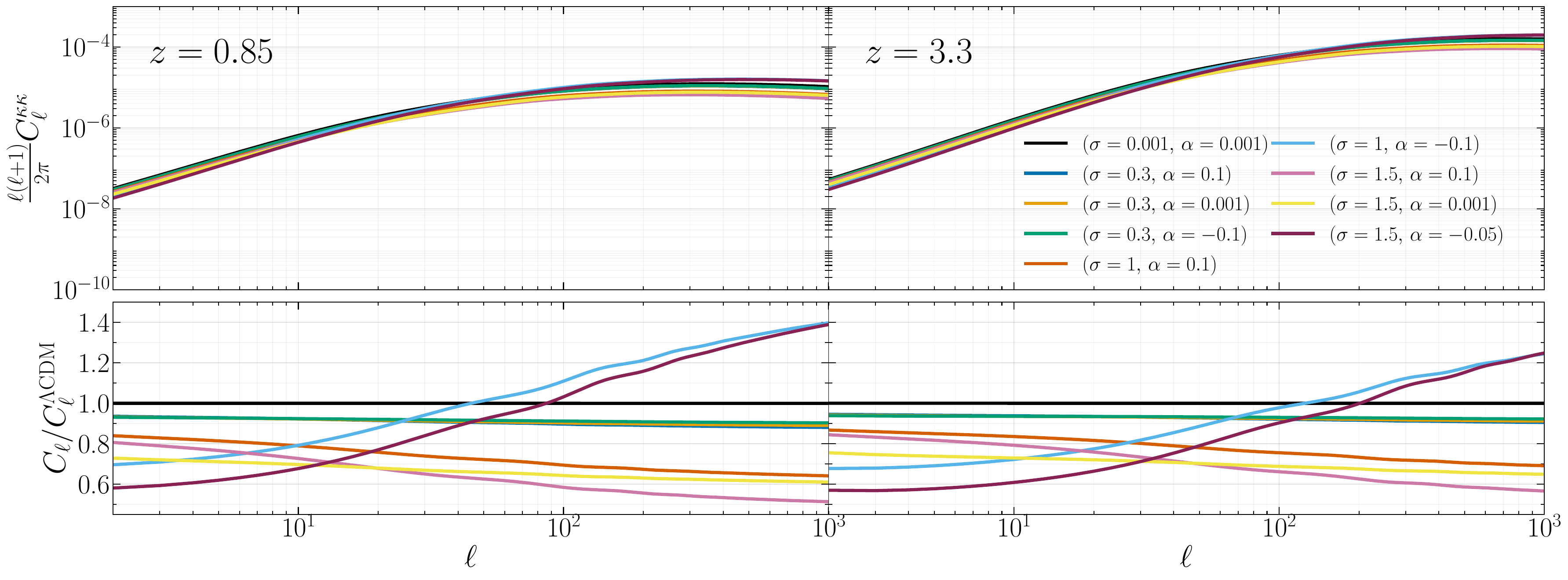}
\caption{
Weak-lensing angular power spectrum for source planes at
$z_s = 0.85$ and $z_s = 3.3$ for representative model parameters. The upper
panels show the convergence spectrum, while the lower panels display
the ratio relative to the corresponding $\Lambda$CDM prediction.
}
\label{fig:lensing_cls}
\end{figure}

\begin{figure}[ht]
\centering
\includegraphics[scale=0.38]{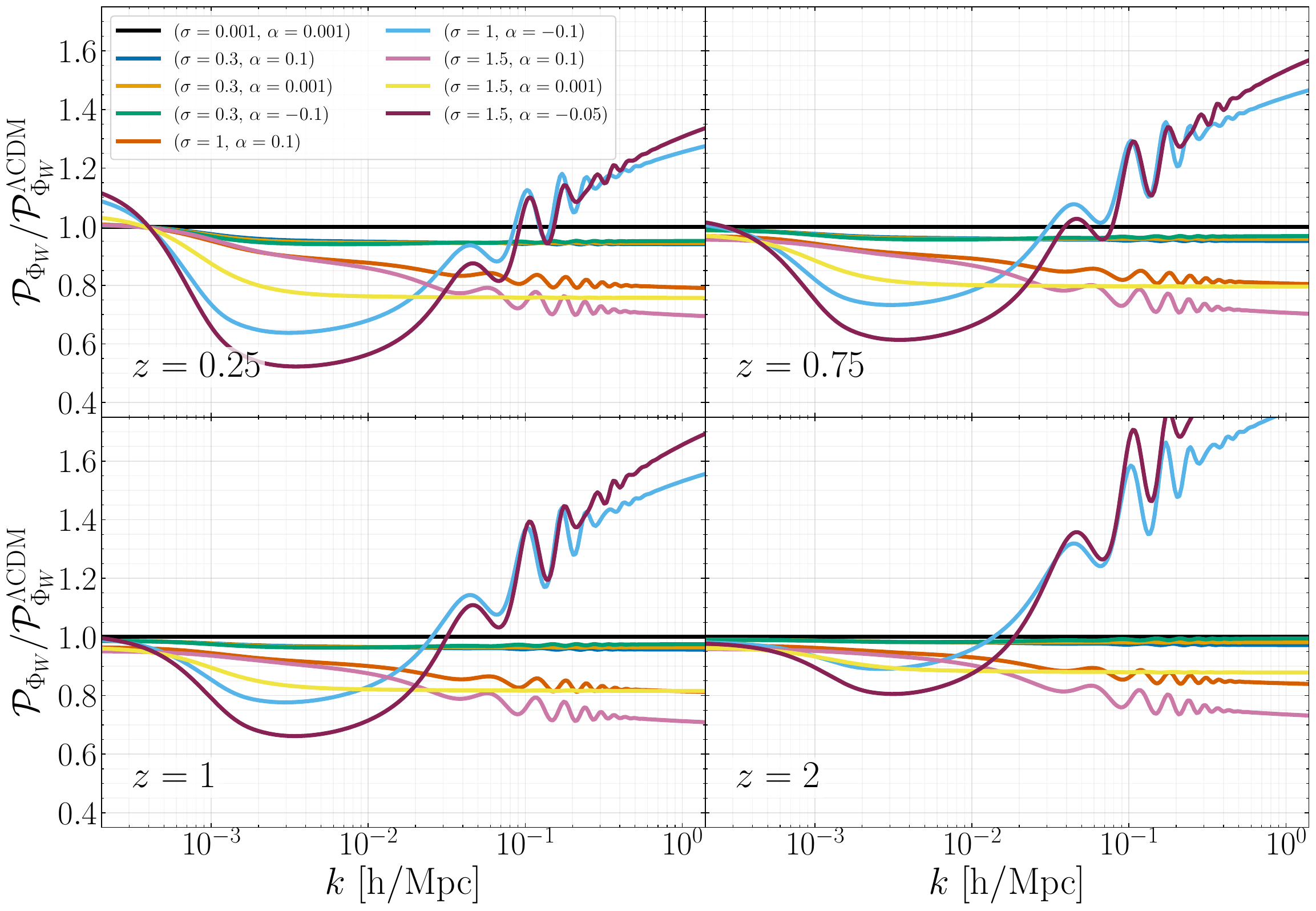}
\caption{
Ratio of the dimensionless Weyl-potential power spectrum
$\mathcal{P}_{\Phi_{\rm W}}$ to the reference $\Lambda$CDM spectrum at
different redshifts. The panels show $z=0.25$, $0.75$, $1$, and $2$ for the
same parameter choices used in Fig.~\ref{fig:lensing_cls}. Although some
models can enhance the Weyl power on the largest scales at late times, the
power is generally suppressed over a broad range of large and intermediate
scales at the redshifts that contribute significantly to the lensing kernel.
This redshift dependence helps explain why the projected convergence spectrum
can be suppressed even when the equal-time Weyl spectrum today is not
uniformly below the $\Lambda$CDM model.
}
\label{fig:Weyl_spectra}
\end{figure}

\subsubsection{The late Integrated Sachs--Wolfe effect}
\label{subsubsec:isw}

The late Integrated Sachs--Wolfe (ISW) effect provides another direct probe
of the metric sector. Unlike weak lensing, which is sensitive to the
line-of-sight projection of the Weyl potential itself, the ISW effect is
sourced by the time variation of the metric potentials along the photon
trajectory \citep{1967ApJ...147...73S}. 

Here we focus on the late-time ISW contribution, rather than on the full
CMB ISW signal integrated from the last-scattering surface. This separation
is possible because the ISW contribution is a line-of-sight integral and is
therefore additive. The full integral can be split into contributions from
different epochs. The early ISW contribution is generated mainly around the
radiation--matter transition, when the potentials are evolving. By
contrast, during matter domination with negligible
anisotropic stress, the linear Weyl potential is
approximately time independent, so the ISW source is strongly suppressed.
The late ISW contribution is then generated at low redshift, when dark
energy or any other late-time dynamics causes the metric
potentials to evolve.

The ISW part of the photon geodesic equation can be expressed as a
perturbation to the observed redshift. Writing the observed redshift as
$z=\bar z+\delta z$, where $\bar z$ is the background redshift and
$\delta z$ is the perturbation induced by metric fluctuations along the
line of sight, the corresponding temperature fluctuation is
$\Theta\equiv \Delta T/\bar T=-\delta z/(1+\bar z)$. Here $\bar T$ denotes
the mean CMB temperature in the unperturbed background, while $\Delta T$ is
the perturbation around this mean. The minus sign follows from
$T\propto (1+z)^{-1}$. Thus a positive perturbation to the observed redshift
corresponds to a lower observed temperature.

Using the same Weyl potential as in
Eq.~\eqref{eq:lenspot_full_dollars2}, the late ISW contribution accumulated
up to a redshift $z_s$, or equivalently up to comoving distance $\chi_s$,
can be written as
\begin{equation}
\Theta_{\rm ISW}(\hat{\mathbf n},z_s)
\equiv
\frac{\Delta T_{\rm ISW}}{\bar T}(\hat{\mathbf n},z_s)
=
-\frac{\delta z_{\rm ISW}}{1+\bar z}
=
\int_0^{\chi_s}{\rm d}\chi\,
\frac{\partial}{\partial\eta}
\left[
\Phi+\Psi
\right]
\bigl(\chi\hat{\mathbf n},\eta_0-\chi\bigr).
\label{eq:isw_temp_chi_phi_psi}
\end{equation}
Equivalently, since $\Phi_{\rm W}=(\Phi+\Psi)/2$,
\begin{equation}
\Theta_{\rm ISW}(\hat{\mathbf n},z_s)
=
2\int_0^{\chi_s}{\rm d}\chi\,
\Phi_{\rm W}'
\bigl(\chi\hat{\mathbf n},\eta_0-\chi\bigr),
\label{eq:isw_temp_chi_weyl}
\end{equation}
Here $\hat{\mathbf n}$ is the observed line-of-sight direction,
$\chi=\eta_0-\eta$, and $\chi_s$ sets the upper limit of the late-time
projection. The full CMB ISW contribution is recovered by taking
$\chi_s=\chi_*$, where $\chi_*$ is the comoving distance to the
last-scattering surface. The finite-$z_s$ expression instead isolates the
part of the ISW signal generated between today and the chosen source redshift.

On large linear scales, the late ISW signal is generated by the time
evolution of the Weyl potential. At late times, the departure from a matter-dominated regime changes both the expansion
rate and the growth rate of perturbations, so that the metric potentials are
no longer constant and $\Phi_{\rm W}'$ becomes non-zero. In the present model, this evolution is further modified by the non-minimal coupling and scalar-field
perturbations, which alter the linear constraint and evolution equations for
$\Phi$ and $\Psi$. As a result, the time derivative of the Weyl potential,
and therefore the late ISW signal, can differ from its $\Lambda$CDM
behaviour \citep{Adamek:2019vko, Beck:2018owr}.

There is also a genuinely non-linear contribution, the non-linear Rees--Sciama effect
\citep{1968Natur.217..511R}, sourced by the time evolution of non-linear
structures. Physically, photons crossing evolving overdensities and voids
do not in general gain and lose exactly compensating amounts of energy,
because the potentials evolve while the photons propagate through them.
This effect is most relevant on non-linear scales. The spectra computed
here with a linear Boltzmann solver do not include this non-linear
Rees--Sciama contribution. They should therefore be interpreted as the
linear late-ISW response of the background and scalar perturbations,
encoded in the evolution of $\Phi_{\rm W}'$.

As for the convergence spectrum in
Eq.~\eqref{eq:Cl_kappa_full_Phi_dollars2}, it is useful to write the ISW
angular spectrum directly in terms of an unequal-time metric power spectrum.
We define the unequal-time power spectrum of the Weyl-potential derivative
by
\begin{equation}
\left\langle
\Phi_{\rm W}'(\mathbf k,\chi)
\Phi_{\rm W}'{}^{\ast}(\mathbf k',\chi')
\right\rangle
=
(2\pi)^3
\delta_{\rm D}(\mathbf k-\mathbf k')\,
P_{\Phi_{\rm W}'}(k;\chi,\chi') ,
\label{eq:isw_derivative_power}
\end{equation}
where
\(\Phi_{\rm W}'(\mathbf k,\chi)\equiv
\Phi_{\rm W}'(\mathbf k,\eta_0-\chi)\). The corresponding full-sky ISW
auto-spectrum truncated at $z_s$ is then
\begin{equation}
C_\ell^{\Theta\Theta} (z_s)
=
\frac{8}{\pi}
\int_0^\infty {\rm d}k\,k^2
\int_0^{\chi_s}{\rm d}\chi
\int_0^{\chi_s}{\rm d}\chi'\,
j_\ell(k\chi)\,
j_\ell(k\chi')\,
P_{\Phi_{\rm W}'}(k;\chi,\chi') .
\label{eq:Cl_ISW_weylprime_full}
\end{equation}
 This form is the direct analogue of the weak-lensing expression in
Eq.~\eqref{eq:Cl_kappa_full_Phi_dollars2}, but with two important
differences. First, the ISW kernel does not contain the geometrical lensing
efficiency $\chi(\chi_s-\chi)/\chi_s$, because the ISW effect is a
temperature shift accumulated along the photon trajectory rather than an
angular distortion of a source plane. Second, the relevant field is not
$\Phi_{\rm W}$ itself but its conformal-time derivative
$\Phi_{\rm W}'$. The ISW signal is therefore especially sensitive to
late-time changes in the metric potentials, while lensing is more directly
sensitive to their projected amplitude.

Equation~\eqref{eq:Cl_ISW_weylprime_full} also explains why the linear ISW
auto-spectrum is mainly a large-scale observable. In the linear regime the
ISW source is the slow time evolution of the metric potentials, not the
rapid motion of matter inside small-scale structures. For non-relativistic
matter the potentials evolve on a cosmological time scale, so one may
schematically write $\Phi_{\rm W}'\sim {\cal H}\Phi_{\rm W}$, with the
proportionality determined by the evolution of the growth rate, the
background expansion, and, in curvature-coupled models, the modified gravitational response.
The scale dependence of $\Phi_{\rm W}'$ therefore largely follows that of
$\Phi_{\rm W}$. Since the linear constraint equations relate the potential
to the density through a Poisson-like relation,
$\Phi_{\rm W}(k)\sim \delta_{\rm m}(k)/k^2$, short-wavelength density modes
generate comparatively weak potential perturbations, and hence a suppressed
linear ISW source.

There is an additional suppression from the projection itself. The ISW
temperature shift is a signed line-of-sight integral of $\Phi_{\rm W}'$.
For modes whose wavelength is short compared with the distance over which
the source evolves, the photon samples many spatial phases along its path,
so the associated redshift shifts partly cancel. In the full-sky expression
this appears as the oscillatory spherical-Bessel kernels in
Eq.~\eqref{eq:Cl_ISW_weylprime_full}. The linear ISW signal is therefore
concentrated on large angular scales, where the evolving potentials remain
coherent over a significant part of the line of sight
\citep{Crittenden:1995ak,Afshordi:2004kz,Ho:2008bz}.

In our analysis we compute the relevant angular spectra using
\texttt{hi\_class}, which solves the modified background and linear
perturbation equations of the curvature-coupled model and performs the line-of-sight
projection numerically. Since the calculation is linear, the ISW spectra
shown below should be interpreted as the linear late-time ISW contribution.
They do not include the non-linear Rees--Sciama correction.

\begin{figure}[ht]
\centering
\includegraphics[scale=0.26]{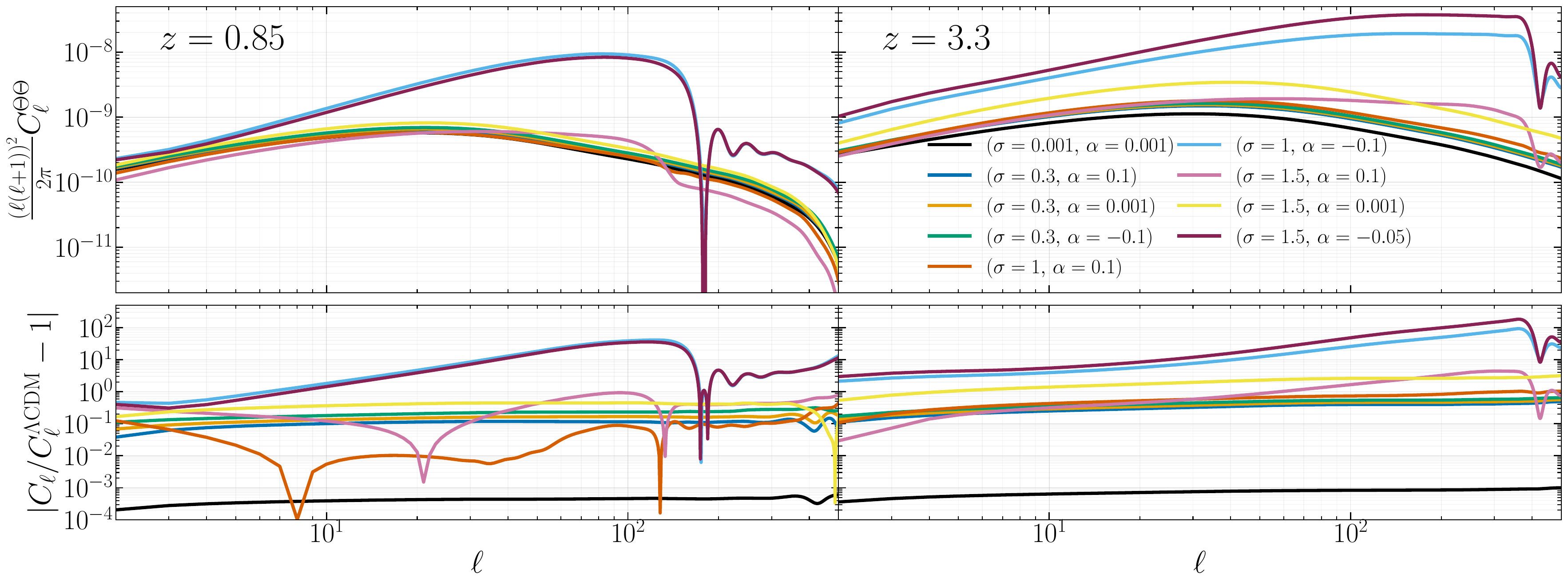}
\caption{
Late Integrated Sachs--Wolfe angular power spectra for truncated
line-of-sight integrals up to $z_s=0.85$ and $z_s=3.3$ in the curvature-coupled
model.
The upper panels show the late-ISW angular power spectra, while the lower
panels show the absolute fractional deviations from the corresponding
$\Lambda$CDM predictions. Sharp dips or broken
segments in the lower panels occur when one of the model spectra crosses the
$\Lambda$CDM prediction. They are therefore artefacts of plotting the
absolute deviation on a logarithmic scale, not physical divergences.
}
\label{fig:isw_cls}
\end{figure}

Figure~\ref{fig:isw_cls} shows the late-ISW angular power spectra for the representative parameter choices, compared with the corresponding $\Lambda$CDM predictions for two
source redshifts, $z_s=0.85$ and $z_s=3.3$. In the upper panels,
the signal is concentrated at low multipoles and is rapidly suppressed
towards smaller angular scales. 
The multipole dependence of this suppression varies across the parameter
space, because the scalar field modifies both the amplitude and the
scale-dependence of the Weyl-potential evolution. For most of the parameter
choices shown here, the spectra are enhanced relative to $\Lambda$CDM,
especially for the deeper integral with $z_s=3.3$. This indicates a larger
integrated time variation of $\Phi_{\rm W}$ along the line of sight.

The two source redshifts highlight that the modifications introduced by the
model do not correspond to a simple overall rescaling of the ISW amplitude. For $z_s=0.85$, the spectra mainly
probe the recent evolution of the metric potentials, while for $z_s=3.3$ the
projection also includes earlier stages of the scalar-field evolution and a
larger range of comoving distances. As a result, the relative ordering of the
models and the multipole dependence of the deviations are different in the
two panels. Some models remain close to $\Lambda$CDM over most of the
multipole range, whereas others become increasingly enhanced at intermediate
and high $\ell$. This is especially clear for the negative-coupling,
large-$\sigma$ cases, $(\sigma,\alpha)=(1,-0.1)$ and $(1.5,-0.05)$, which
produce the largest ISW amplitudes for the deeper projection. These are also
the models for which Fig.~\ref{fig:mu_gamma} shows the largest departures
in the gravitational slip parameter $\gamma=\Phi/\Psi$, indicating a strong
modification of the relative evolution of the two metric potentials and
therefore of the Weyl combination that sources the ISW effect.

A striking feature of these same negative-$\alpha$, large-$\sigma$ models is
the sharp suppression in the upper panels of Fig.~\ref{fig:isw_cls}. For
$z_s=0.85$ this occurs around $\ell\sim 150$, while for $z_s=3.3$ the
corresponding feature is shifted to higher multipoles. This shift is
consistent with a projection effect: a feature associated with a characteristic
physical scale contributes roughly at $\ell\sim k\chi_{\rm eff}$, so extending
the line-of-sight integral to larger redshift moves the same scale to larger
angular multipoles. The suppression should therefore be interpreted as a
scale-dependent feature in the projected ISW transfer function, rather than as
a local disappearance of the ISW source at one redshift.

\begin{figure}[ht]
\centering
\includegraphics[scale=0.25]{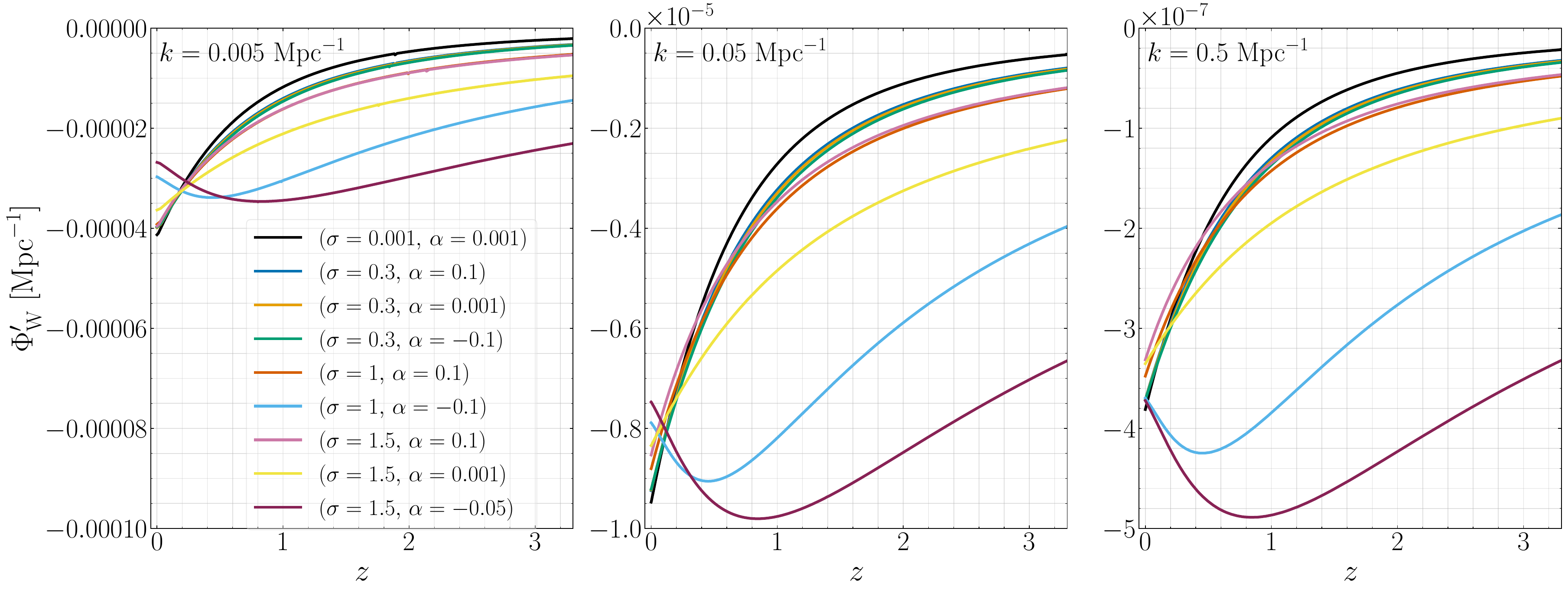}
\caption{Conformal-time derivative of the Weyl potential,
$\Phi_{\rm W}'$, as a function of redshift for three representative Fourier
modes. This quantity is the local source of the linear late-ISW effect.
Negative values correspond to a decay of the Weyl potential in our sign
convention. The different parameter choices modify both the amplitude and the redshift
dependence of this decay, with a scale dependence that becomes especially
visible when comparing the three $k$ modes. At higher redshift the different
models are more clearly separated, while at low redshift several curves
approach or cross the $\Lambda$CDM behaviour,
represented here by the $(\sigma,\alpha)=(0.001,0.001)$ case, indicating a weaker potential
decay than in $\Lambda$CDM over part of the recent evolution. This helps
explain why some models can give a reduced low-redshift ISW contribution for
$z_s=0.85$, while producing a stronger signal when the projection is extended
to $z_s=3.3$, as shown in Fig.~\ref{fig:isw_cls}.
}
\label{fig:weyl_prime_evolution}
\end{figure}

The origin of these features can be understood from
Fig.~\ref{fig:weyl_prime_evolution}, which shows $\Phi_{\rm W}'$ for
representative Fourier modes. The ISW spectrum is a projection of this source
over both time and scale, so it cannot be inferred from a single value of
$\Phi_{\rm W}'(k,z)$. Nevertheless, the negative-$\alpha$, large-$\sigma$
models show a clearly non-monotonic evolution: the magnitude of
$\Phi_{\rm W}'$ first increases and then decreases, with the turnover occurring
at a wavenumber-dependent redshift. When such a source is projected with the
oscillatory spherical-Bessel kernels, different radial ranges can contribute
with opposite phases and partially cancel for particular multipoles. This
naturally produces a sharp suppression in the projected ISW transfer function,
and hence in $C_\ell^{\Theta\Theta}$. The small oscillations visible after the
drop in Fig.~\ref{fig:isw_cls}, especially for $z_s=0.85$, can be understood as the neighbouring
multipoles moving through alternating regions of constructive and destructive
projection. They are more apparent for the lower-redshift integral because the
shorter line of sight provides less radial averaging of the scale-dependent
structure in $\Phi_{\rm W}'$. By contrast, cases with a smoother and more
monotonic evolution of $\Phi_{\rm W}'$ lead to smoother ISW spectra.

Finally, the sharp dips or broken segments in the lower ratio panels of
Fig.~\ref{fig:isw_cls} should be interpreted with some care. In several
cases they occur when a model spectrum crosses the corresponding
$\Lambda$CDM prediction, so that the absolute fractional deviation becomes
very small and produces a sharp feature on a logarithmic scale. The crossing
itself can be physical, since it reflects a multipole at which the projected
ISW power in the curvature-coupled model matches the $\Lambda$CDM value. However, the
apparent depth or discontinuity of the feature is partly a consequence of
plotting $|C_\ell/C_\ell^{\Lambda{\rm CDM}}-1|$ on a logarithmic axis.

\section{Conclusions}
\label{sec:conclusions}

In this work, we have presented a comprehensive study of the background and
linear cosmological predictions of a curvature-coupled dark energy model with the
shifted quartic non-minimal coupling
$f(\varphi)=\alpha(\varphi^2-\varphi_{\rm today}^2)^2$
and the inverse power-law potential
$V(\varphi)=\Lambda\varphi^{-\sigma}$. We implemented the model in a modified
version of \texttt{hi\_class}, enabling a self-consistent computation of its
background evolution and linear perturbations. Without requiring an additional screening mechanism to satisfy the local constraints considered here, the shifted coupling preserves the present-day normalization of gravity and satisfies the tight bounds on the effective
gravitational coupling and post-Newtonian parameters, while allowing the scalar field to remain cosmologically active at early times and to drive late-time acceleration.

The model exhibits a phenomenology that is not available in minimally coupled
canonical quintessence. In particular, parts of the parameter space undergo a
smooth effective crossing of the phantom divide, generated by the
non-minimal coupling terms in the effective scalar fluid description without
introducing instabilities. The non-minimal coupling also
modifies the expansion history both before recombination and at late times.
For negative values of $\alpha$, the early-time scalar contribution can
increase the pre-recombination expansion rate and reduce the sound horizon,
shifting the acoustic scale in the direction associated with a higher inferred
value of $H_0$ and potentially alleviating the Hubble tension.

At the perturbation level, the model produces correlated but scale-dependent changes
in the effective Newtonian response, the gravitational slip, matter
clustering, and the Weyl potential. These modifications propagate differently
into the observables considered here. We find sizeable departures from the
$\Lambda$CDM model in the matter and metric-potential spectra, suppressions of
the weak-lensing convergence spectrum at low multipoles, followed in some
negative-$\alpha$ models by an enhancement at higher multipoles, and
substantial changes in the late Integrated Sachs--Wolfe signal. The distinct scale and redshift dependence of these observables provides complementary ways of testing the model and shows that its phenomenology cannot be reduced
to a simple rescaling of the gravitational coupling or the expansion rate.

The next stage of this work will focus on two main directions. First, we plan
to perform a full Bayesian parameter inference to constrain the model parameters and investigate whether the
best-fitting solutions can simultaneously alleviate the Hubble tension and
exhibit the effective phantom-crossing behaviour suggested by recent
observational analyses, while remaining consistent with the full set of
cosmological data. 
Second, we plan to extend the model
to the non-linear regime through dedicated $N$-body simulations. This will enable
a self-consistent study of structure formation beyond linear theory, test whether the curvature-induced, environment-dependent scalar dynamics
leads to an efficient chameleon-like suppression in high-density environments, and provide accurate predictions for
non-linear observables, for which the standard
non-linear prescriptions currently implemented in \texttt{hi\_class} are not
applicable to the present model.

\section{Acknowledgments}
\label{sec:acknowledgements}
We would like to thank Julian Adamek and Martin Kunz for helpful discussions. We thank the Research Council of Norway for their support and the resources provided by UNINETT Sigma2 -- the National Infrastructure for High-Performance Computing and Data Storage in Norway. OpenAI ChatGPT was used for language refinement,
clarity improvements, and consistency checks of some of the author-provided derivations.

\section*{Data and code availability}
The codes used to generate the numerical data, together with the Jupyter
notebooks used to reproduce the figures in this work and the modified version
of \texttt{hi\_class} implementing the curvature-coupled dark energy model,
are publicly available at
\href{https://github.com/abdolalibanihashemi/hi_class_CCDE}
{\texttt{hi\_class\_CCDE}} GitHub repository.

\appendix
\section*{Appendix}
\section{Linear perturbation equations}
\label{app:pert_eqs}
In this appendix we collect the linear perturbation equations of the
curvature-coupled model written explicitly in terms of the physical scalar-field
perturbation $\delta\varphi$. They are obtained by varying the covariant
action with respect to the metric and the scalar field, expanding around a
spatially flat FLRW background, and keeping only terms linear in the
perturbations. These equations constitute the linear perturbation equations discussed in Sec.~\ref{sec:linear_pert} and provide the
starting point for future non-linear implementations of the model.

We work in the Poisson gauge and write the perturbed metric as
\begin{equation}
ds^2 = a^2(\eta)\left[-(1+2\Psi)\,d\eta^2 + (1-2\Phi)\,\delta_{ij}\,dx^i dx^j\right],
\label{eq:pertmetric_app}
\end{equation}
where $\Phi$ and $\Psi$ are the Bardeen potentials. The scalar field is
decomposed as
\begin{equation}
\varphi(\eta,\mathbf{x})=\varphi(\eta)+\delta\varphi(\eta,\mathbf{x}),
\end{equation}
and $\mathcal H \equiv a'/a$, with a prime denoting differentiation with
respect to conformal time .

The linearized equations can then be organized into the $\eta\eta$
constraint, the traceless $ij$ equation, the $\eta i$ momentum equation,
and the perturbed Klein--Gordon equation. No quasistatic approximation has been made, and all metric and scalar-field
time derivatives are kept. The equations are therefore valid on all linear scales. Importantly, this formulation contains no division by $\varphi'$ and therefore
remains regular at the isolated turning points where $\varphi'=0$, unlike
the standard \texttt{hi\_class} variable
$V_X=a\,\delta\varphi/\varphi'$.

\paragraph{$\eta\eta$ equation.}
The time-time component of the modified Einstein equations gives
\begin{align}
&(1+f)\,\Delta\Phi
-3\left[\mathcal H(1+f)+\frac{1}{2}\varphi' f_{,\varphi}\right]\Phi'
-3\left[\mathcal H^2(1+f)-\frac{\varphi'^2}{6M_{\rm P}^2}
      +\mathcal H \varphi' f_{,\varphi}\right]\Psi
\nonumber\\
&
+\frac{1}{2}\left(3\mathcal H f_{,\varphi}-\frac{\varphi'}{M_{\rm P}^2}\right)
\delta\varphi'
+\frac{1}{2}\left\{
3\mathcal H\left(\mathcal H f_{,\varphi}+\varphi' f_{,\varphi\varphi}\right)
-\frac{a^2}{M_{\rm P}^2}V_{,\varphi}
\right\}\delta\varphi
-\frac{1}{2}f_{,\varphi}\,\Delta\delta\varphi
=
\frac{a^2}{2M_{\rm P}^2}\,\delta\rho \; .
\label{eq:tt_pert_eq_app}
\end{align}

\paragraph{Traceless $ij$ equation.}
The traceless spatial part yields
\begin{align}
\left(\delta^i_k\delta^j_l-\frac{1}{3}\delta^{ij}\delta_{kl}\right)
\left[
\partial_i\partial_j(\Phi-\Psi)
-\frac{f_{,\varphi}}{1+f}\,\partial_i\partial_j\delta\varphi
\right]
=
\frac{a^2}{M_{\rm P}^2(1+f)}
\left(
\delta_{ki}T^i_{\ l}-\frac{1}{3}\delta_{kl}T^i_{\ i}
\right)\; .
\label{eq:ij_traceless_app}
\end{align}
This equation makes explicit that the non-minimal coupling sources a
gravitational slip even in the absence of intrinsic matter anisotropic
stress.

\paragraph{$\eta i$ equation.}
The momentum constraint reads
\begin{align}
&-\left[\mathcal H(1+f)+\frac{1}{2}\varphi' f_{,\varphi}\right]\partial_i\Psi
-(1+f)\,\partial_i\Phi'
-\frac{1}{2}\left[
\mathcal H f_{,\varphi}
-\left(\frac{1}{M_{\rm P}^2}+f_{,\varphi\varphi}\right)\varphi'
\right]\partial_i\delta\varphi
\nonumber\\
& +\frac{1}{2}f_{,\varphi}\,\partial_i\delta\varphi'
=
\frac{a^2}{2M_{\rm P}^2}\,\delta T^0_{\ i} \; .
\label{eq:ti_pert_eq_app}
\end{align}

\paragraph{Perturbed Klein--Gordon equation.}
Varying the action with respect to the scalar field and linearizing gives
\begin{align}
& \delta\varphi''
+2\mathcal H\,\delta\varphi'
-\Delta\delta\varphi
-\left[
3M_{\rm P}^2 f_{,\varphi\varphi}\left(\mathcal H^2+\mathcal H'\right)
-a^2 V_{,\varphi\varphi}
\right]\delta\varphi
\nonumber\\
&
+\left(3M_{\rm P}^2 f_{,\varphi}\mathcal H-\varphi'\right)
\left(\Psi'+3\Phi'\right)
+2a^2 V_{,\varphi}\Psi
-2M_{\rm P}^2 f_{,\varphi}\Delta\Phi
+M_{\rm P}^2 f_{,\varphi}\Delta\Psi
+3M_{\rm P}^2 f_{,\varphi}\Phi''
=0 \,.
\label{eq:kg_pert_eq_app}
\end{align}

Eqs.~\eqref{eq:tt_pert_eq_app}--\eqref{eq:kg_pert_eq_app} form the
linear system governing the coupled evolution of matter, metric, and
scalar-field perturbations in the curvature-coupled model. In practice, these equations
are solved numerically in \texttt{hi\_class} through the equivalent
Horndeski formulation discussed in Appendix~\ref{app:hiclass}

\section{Implementation in \texttt{hi\_class}}
\label{app:hiclass}
$G_i(\varphi,X)$ functions are the only theoretical ingredients required to implement the model in \texttt{hi\_class}.
 The code then self-consistently computes the background evolution and evaluates the effective Planck mass, the EFT
$\alpha$-functions, and the coefficients entering the linear equations.

The implementation consists of two independent steps. First, the dimensionless Horndeski functions $\tilde G_i$ and their derivatives are
provided to the code following the normalization convention of
\citep{Bellini:2019syt}. Second, the standard scalar perturbation
evolution based on the auxiliary variable
$V_X=a\,\delta\varphi/\varphi'$ is replaced by the direct evolution of the
physical scalar-field perturbation $\delta\varphi$, as discussed in
Sec.~\ref{sec:linear_pert}. Apart from this modification, the remaining
background, Einstein, matter, and Boltzmann evolution are unchanged.

In order to implement the $G_i(\varphi,X)$ functions in
\texttt{hi\_class}, they must first be written in the dimensionless form
adopted by the code, following the procedure described in
\cite{Bellini:2019syt}. We denote dimensionless quantities with a tilde,
such that $\tilde{\varphi}=\varphi/M_{\rm P}$ and
$\widetilde{\partial_\mu\varphi}=L\,\partial_\mu\tilde{\varphi}$, where
$L\equiv1\,{\rm Mpc}$ is the fixed space-time length scale used internally
by \texttt{hi\_class}. The dimensionless kinetic term is therefore
$\tilde{X}=L^2X/M_{\rm P}^2$.
When expressing the model functions in terms of the dimensionless field
$\tilde{\varphi}$, the corresponding normalization factors are absorbed into the model parameters. In particular, the coupling parameter \eqref{eq:DE_coupling2} becomes
$\tilde{\alpha}\equiv\alpha M_{\rm P}^4$, while the potential amplitude \eqref{eq:DE_potential2}
entering the dimensionless implementation is
$ \tilde\Lambda \equiv
\frac{\Lambda}{H_0^2 M_{\rm P}^{2+\sigma}}.$
The dimensionless scalar-field potential is therefore
\be
\tilde V(\tilde{\varphi})
=
(H_0L)^2 \tilde\Lambda\,\tilde{\varphi}^{-\sigma},
\ee
making it dimensionless while preserving the normalization appropriate
for late-time dark energy. The coupling function expressed in terms of the dimensionless field becomes
\begin{equation}
\tilde f(\tilde{\varphi})
\equiv
f(M_{\rm P}\tilde{\varphi})
=
\tilde{\alpha}
\left(
\tilde{\varphi}^2-\tilde{\varphi}_{\rm today}^2
\right)^2 .
\end{equation}
Since \texttt{hi\_class} adopts units in which $M_{\rm P}=1$ and $H_0$ has
units of ${\rm Mpc}^{-1}$, we implement the following functions and all their required derivatives with respect to $\tilde{\varphi}$ and $\tilde{X}$:

\begin{align}
\tilde{G}_2[\tilde{\varphi}, \tilde{X}] &= \tilde{X} -  \tilde V(\tilde{\varphi})\,,\\
\tilde{G}_4[\tilde{\varphi}] &= \frac{1}{2} \left[1 + \tilde f(\tilde{\varphi})\right]\,.
\end{align}
Since $L=1\,{\rm Mpc}$ is fixed internally, the factor $(H_0L)^2$
appearing above is represented numerically in the code simply by $H_0^2$. Moreover, because $M_{\rm P}=1$ internally, $\tilde{\alpha}$ has the same numerical
value as the parameter $\alpha$ quoted in the main text, while $\tilde{\Lambda}$ is the dimensionless potential amplitude in
the numerical implementation.
\section{Details of the weak-lensing formalism}
\label{app:lensing_details}

This appendix collects the geometrical and spectral details used in
Sec.~\ref{subsubsec:lensing}. The main text gives the lensing observables
used in our analysis, while here we provide the derivation of the line-of-sight
kernel, the sign conventions for convergence and shear, the relation between
shear and the scalar lensing potential, and the Limber limit of the full-sky
projection.

\subsection{Light-cone geometry and the deflection kernel}
\label{app:lensing_geometry}

We first describe the geometrical ingredients entering the deflection
angle in Eq.~\eqref{eq:deflection_dollars2}. In the Born approximation, the
metric potentials are evaluated along the unperturbed background photon
trajectory,
\begin{equation}
\mathbf x(\chi)=\chi\,\hat{\mathbf n},
\qquad
\eta(\chi)=\eta_0-\chi .
\end{equation}
The second relation follows from the radial null condition in the
background FLRW spacetime,
\begin{equation}
ds^2=a^2(\eta)(-d\eta^2+d\chi^2)=0,
\end{equation}
which gives $d\eta=-d\chi$ along the observer's past light cone. Thus
$\bigl(\chi\hat{\mathbf n},\eta_0-\chi\bigr)$ denotes the spacetime point
on the unperturbed past light cone at comoving distance $\chi$ and observed
direction $\hat{\mathbf n}$.

The gradient appearing in Eq.~\eqref{eq:deflection_dollars2} is an angular
gradient at fixed comoving distance. In an orthonormal basis tangent to the
unit two-sphere,
\begin{equation}
\nabla_{\!\hat{\mathbf n}}
=
\mathbf e_\theta\,\partial_\theta
+
\mathbf e_\phi\,\frac{1}{\sin\theta}\partial_\phi .
\end{equation}
The corresponding physical transverse gradient at distance $\chi$ is
\begin{equation}
\nabla_\perp
=
\frac{1}{\chi}\nabla_{\!\hat{\mathbf n}} .
\end{equation}
This relation explains the factor $1/\chi$ multiplying
$\nabla_{\!\hat{\mathbf n}}\Phi_{\rm W}$ in
Eq.~\eqref{eq:deflection_dollars2}: the physical deflection is sourced by
the transverse gradient of the Weyl potential along the photon path.

The remaining factor in the deflection kernel has a simple geometrical
origin. Consider a perturbation localized at a lens plane with comoving
distance $\chi$, between the observer and the source plane at $\chi_s$. If
the ray is deflected by a small angle $\delta\boldsymbol{\theta}$ at this
lens plane, then its intersection with the source plane is shifted
transversely by
\begin{equation}
\delta\mathbf x_s
\simeq
(\chi_s-\chi)\,\delta\boldsymbol{\theta}.
\end{equation}
The corresponding angular shift on the observer's sky is
\begin{equation}
\boldsymbol{\alpha}
\simeq
\frac{\delta\mathbf x_s}{\chi_s}
\simeq
\frac{\chi_s-\chi}{\chi_s}
\delta\boldsymbol{\theta}.
\end{equation}
Therefore perturbations very close to the source plane are geometrically
suppressed because $\chi_s-\chi\to0$. Combining this source-plane projection factor with the transverse-gradient
relation above gives the geometrical structure of the deflection kernel in
Eq.~\eqref{eq:deflection_dollars2}.

Near the observer, the apparent $1/\chi$ factor in
Eq.~\eqref{eq:deflection_dollars2}, or equivalently in
Eq.~\eqref{eq:lenspot_full_dollars2}, should not be interpreted as a
physical enhancement. At fixed $\chi$, an angular separation
$\delta\theta$ corresponds to a transverse separation
$\delta x_\perp\simeq\chi\delta\theta$, so
\begin{equation}
\nabla_{\!\hat{\mathbf n}}
=
\chi\nabla_\perp .
\end{equation}
Consequently,
\begin{equation}
\frac{1}{\chi}
\nabla_{\!\hat{\mathbf n}}
\Phi_{\rm W}\bigl(\chi\hat{\mathbf n},\eta_0-\chi\bigr)
=
\nabla_\perp
\Phi_{\rm W}\bigl(\chi\hat{\mathbf n},\eta_0-\chi\bigr),
\end{equation}
which is finite as $\chi\to0$ provided the physical transverse gradient of
the potential is regular at the observer.

\subsection{Distortion tensor and geometrical lensing efficiency}
\label{app:lensing_distortion}

We now derive the decomposition of the angular distortion tensor mentioned in Sec.~\ref{subsubsec:lensing}, and show how the geometrical lensing
efficiency appearing in the main text arises from the projected convergence
kernel. The absolute deflection angle is not itself directly observable,
since a nearly uniform shift of all source positions can be absorbed into a
redefinition of the apparent sky coordinates. The observable weak-lensing
information is instead contained in how the deflection varies across
neighbouring light rays. This variation is described by the angular
distortion tensor
\begin{equation}
\mathcal D_{ab}
\equiv
\nabla_a\alpha_b
=
\nabla_a\nabla_b\phi^{\rm L} ,
\end{equation}
where $a,b$ are indices tangent to the unit two-sphere and $\nabla_a$ is the
angular covariant derivative. Thus $\mathcal D_{ab}$ is the angular Hessian
of the lensing potential.

At linear order in scalar perturbations, $\mathcal D_{ab}$ is symmetric and
contains no image rotation. It can therefore be decomposed into an isotropic
trace part and an anisotropic trace-free part. Let $q_{ab}$ be the metric on
the unit two-sphere,
\begin{equation}
d\Omega^2
=
q_{ab}\,d\theta^a d\theta^b
=
d\theta^2+\sin^2\theta\,d\phi^2 .
\end{equation}
The trace of the distortion tensor is
\begin{equation}
\mathcal D
\equiv
q^{ab}\mathcal D_{ab}.
\end{equation}
Since the screen is two-dimensional, the isotropic part is
\begin{equation}
\mathcal D^{\rm iso}_{ab}
=
\frac{1}{2}\mathcal D\,q_{ab}.
\end{equation}
With the standard weak-lensing image-to-source sign convention, the
convergence is defined by
\begin{equation}
\kappa
\equiv
-\frac{1}{2}\mathcal D
=
-\frac{1}{2}q^{ab}\mathcal D_{ab}.
\end{equation}
The shear tensor is defined as minus the trace-free part of the distortion
tensor,
\begin{equation}
\mathcal E_{ab}
\equiv
-\left(
\mathcal D_{ab}
-
\frac{1}{2}q_{ab}q^{cd}\mathcal D_{cd}
\right).
\end{equation}
We denote the shear tensor by $\mathcal E_{ab}$ rather than by the more usual
symbols $\gamma_{ab}$ or $\sigma_{ab}$, since these symbols are used for
other quantities in this paper. With the above definitions,
\begin{equation}
\mathcal D_{ab}
=
-\kappa q_{ab}
-
\mathcal E_{ab},
\qquad
q^{ab}\mathcal E_{ab}=0 .
\end{equation}
Thus $\kappa$ describes the isotropic focusing of a light bundle, while
$\mathcal E_{ab}$ describes the anisotropic stretching of the image.

To make the physical meaning of the two shear components explicit, it is
useful to work in a local orthonormal screen basis. In such a basis the two
screen directions are unit and mutually orthogonal, so that
$q_{ab}=\delta_{ab}$. This removes coordinate factors, such as the
$\sin^2\theta$ appearing in the spherical coordinate basis, and makes the
physical stretching and compression directly visible. The trace-free shear
tensor can then be written as
\begin{equation}
\mathcal E_{ab}
=
\begin{pmatrix}
\mathcal E_1 & \mathcal E_2 \\
\mathcal E_2 & -\mathcal E_1
\end{pmatrix},
\label{eq:shear_matrix}
\end{equation}
and the distortion tensor becomes
\begin{equation}
\mathcal D_{ab}
=
\begin{pmatrix}
-\kappa-\mathcal E_1 & -\mathcal E_2 \\
-\mathcal E_2 & -\kappa+\mathcal E_1
\end{pmatrix}.
\end{equation}

The sign convention can be checked from the lensing Jacobian. With the
image-to-source convention, the Jacobian maps an observed angular separation
$\delta\theta^b$ to the corresponding source-plane separation
$\delta\beta_a$,
\begin{equation}
\delta\beta_a
=
\mathcal A_{ab}\,\delta\theta^b,
\qquad
\mathcal A_{ab}
=
\delta_{ab}
+
\mathcal D_{ab}.
\end{equation}
Therefore
\begin{equation}
\mathcal A_{ab}
=
\begin{pmatrix}
1-\kappa-\mathcal E_1 & -\mathcal E_2 \\
-\mathcal E_2 & 1-\kappa+\mathcal E_1
\end{pmatrix}.
\end{equation}
For pure convergence,
\begin{equation}
\delta\beta_a
=
(1-\kappa)\delta\theta_a .
\end{equation}
Thus, for $\kappa>0$, a fixed observed angular separation corresponds to a
smaller intrinsic source-plane separation. Equivalently, inverting the map
gives
\begin{equation}
\delta\theta_a
=
\frac{1}{1-\kappa}\delta\beta_a
\simeq
(1+\kappa)\delta\beta_a ,
\end{equation}
so positive convergence magnifies the observed image. The same
image-to-source convention fixes the signs of the shear terms.

Having fixed the decomposition and sign convention of the distortion tensor,
we can now recover the relations used in Sec.~\ref{subsubsec:isw}. Since
$\mathcal D_{ab}=\nabla_a\nabla_b\phi^{\rm L}$, taking the trace gives
\begin{equation}
\label{eq:kappa_lenspot_dollars2}
\kappa(\hat{\mathbf n},z_s)
=
-\frac{1}{2}
q^{ab}\nabla_a\nabla_b
\phi^{\rm L}(\hat{\mathbf n},z_s)
=
-\frac{1}{2}
\nabla_{\!\hat{\mathbf n}}^2
\phi^{\rm L}(\hat{\mathbf n},z_s),
\end{equation}
while taking minus the trace-free part gives
\begin{equation}
\label{eq:shear_lenspot_dollars2}
\mathcal E_{ab}(\hat{\mathbf n},z_s)
=
-\left(
\nabla_a\nabla_b
-
\frac{1}{2}q_{ab}
\nabla_{\!\hat{\mathbf n}}^2
\right)
\phi^{\rm L}(\hat{\mathbf n},z_s).
\end{equation}
Substituting the line-of-sight expression for the lensing potential,
Eq.~\eqref{eq:lenspot_full_dollars2}, into the convergence definition gives
\begin{equation}
\label{eq:kappa_weyl_projection_dollars2}
\kappa(\hat{\mathbf n},z_s)
=
\int_0^{\chi_s}\mathrm d\chi\,
\frac{\chi_s-\chi}{\chi_s\,\chi}\,
\nabla_{\!\hat{\mathbf n}}^2
\Phi_{\rm W}
\bigl(\chi\hat{\mathbf n},\eta_0-\chi\bigr).
\end{equation}
This expression shows explicitly that the convergence is sourced by angular
variations of the Weyl potential along the line of sight.

To connect this full-sky expression with the usual geometrical lensing
efficiency, we take the small-angle, or flat-sky, limit. In this limit, a
small patch of the sky is treated as locally flat, and the angular Laplacian
is related to the physical transverse Laplacian by
\begin{equation}
\nabla_{\!\hat{\mathbf n}}^2
\simeq
\chi^2\nabla_\perp^2 .
\end{equation}
This approximation is valid for small angular separations, or equivalently
at sufficiently large multipoles. Therefore
\begin{equation}
\kappa(\hat{\mathbf n},z_s)
\simeq
\int_0^{\chi_s}\mathrm d\chi\,
\frac{\chi(\chi_s-\chi)}{\chi_s}\,
\nabla_\perp^2
\Phi_{\rm W}
\bigl(\chi\hat{\mathbf n},\eta_0-\chi\bigr).
\end{equation}
The purely geometrical part of the convergence and shear kernel is then
\begin{equation}
W_{\rm geom}(\chi)
=
\frac{\chi(\chi_s-\chi)}{\chi_s}.
\end{equation}
This kernel vanishes at both endpoints, $\chi=0$ and $\chi=\chi_s$, and
reaches its maximum halfway between the observer and the source,
at $\chi=\chi_s/2$.

This statement refers only to the geometrical part of the lensing
efficiency. The observed lensing signal is not determined by
$W_{\rm geom}$ alone, but also by the Weyl potential evaluated along the
photon path. In the angular power spectrum, the same radial kernel weights
unequal-time Weyl-potential correlations at different distances, so the
final result depends on both the source--lens geometry and the scale- and
redshift-dependent evolution of the metric perturbations.

\subsection{Full-sky projection and the Limber limit}
\label{app:lensing_limber}

We now derive the spectral form of the convergence power spectrum used in
Eq.~\eqref{eq:Cl_kappa_full_Phi_dollars2}. The full-sky expression is presented
in Sec.~\ref{subsubsec:lensing}, while here we show how it follows from the line-of-sight
projection of the Weyl potential and from the two-point function
$\langle \kappa_{\ell m}\kappa^\ast_{\ell' m'}\rangle$. This derivation also
makes clear why spherical Bessel functions appear and why the relevant
three-dimensional input is the unequal-time Weyl-potential power spectrum.

We start from the lensing-potential projection,
Eq.~\eqref{eq:lenspot_full_dollars2},
\begin{equation}
\phi^{\rm L}(\hat{\mathbf n},z_s)
=
-2\int_0^{\chi_s}\mathrm d\chi\,
\frac{\chi_s-\chi}{\chi_s\,\chi}\,
\Phi_{\rm W}(\chi\hat{\mathbf n},\eta_0-\chi),
\end{equation}
and from the harmonic-space relation between convergence and lensing
potential,
\begin{equation}
\kappa_{\ell m}(z_s)
=
\frac{1}{2}\ell(\ell+1)\phi^{\rm L}_{\ell m}(z_s).
\end{equation}
It is useful to define the radial weight
\begin{equation}
W_{\phi^{\rm L}}(\chi)
\equiv
\frac{\chi_s-\chi}{\chi_s\,\chi}.
\end{equation}
With this definition,
\begin{equation}
\kappa_{\ell m}(z_s)
=
-\ell(\ell+1)
\int_0^{\chi_s}\mathrm d\chi\,
W_{\phi^{\rm L}}(\chi)\,
\Phi_{{\rm W},\ell m}(\chi),
\end{equation}
where
\begin{equation}
\Phi_{{\rm W},\ell m}(\chi)
\equiv
\int d\Omega_{\hat{\mathbf n}}\,
Y_{\ell m}^{\ast}(\hat{\mathbf n})\,
\Phi_{\rm W}(\chi\hat{\mathbf n},\eta_0-\chi)
\end{equation}
is the angular harmonic coefficient of the Weyl potential on the spherical
shell at radius $\chi$.

To relate this projected field to the three-dimensional Weyl-potential power
spectrum, we Fourier expand
\begin{equation}
\Phi_{\rm W}(\mathbf x,\chi)
=
\int\frac{\mathrm d^3 k}{(2\pi)^3}\,
\Phi_{\rm W}(\mathbf k,\chi)\,
e^{i\mathbf k\cdot\mathbf x},
\qquad
\Phi_{\rm W}(\mathbf k,\chi)
\equiv
\Phi_{\rm W}(\mathbf k,\eta_0-\chi).
\end{equation}
The plane wave is expanded in spherical waves as
\begin{equation}
e^{i\mathbf k\cdot\chi\hat{\mathbf n}}
=
4\pi
\sum_{\ell m}
i^\ell
j_\ell(k\chi)
Y_{\ell m}(\hat{\mathbf n})
Y_{\ell m}^{\ast}(\hat{\mathbf k}).
\end{equation}
The spherical Bessel functions therefore appear because a three-dimensional
Fourier mode is being projected onto angular harmonics on a sphere. Using
this expansion gives
\begin{equation}
\Phi_{{\rm W},\ell m}(\chi)
=
4\pi i^\ell
\int\frac{\mathrm d^3 k}{(2\pi)^3}\,
\Phi_{\rm W}(\mathbf k,\chi)\,
j_\ell(k\chi)
Y_{\ell m}^{\ast}(\hat{\mathbf k}).
\end{equation}

The unequal-time Weyl-potential power spectrum is defined by
\begin{equation}
\left\langle
\Phi_{\rm W}(\mathbf k,\chi)
\Phi_{\rm W}^{\ast}(\mathbf k',\chi')
\right\rangle
=
(2\pi)^3
\delta_{\rm D}(\mathbf k-\mathbf k')\,
P_{\Phi_{\rm W}}(k;\chi,\chi') .
\end{equation}
The term ``unequal-time'' means that the two Weyl potentials are evaluated
at different radial distances, $\chi$ and $\chi'$, or equivalently at
different conformal times on the past light cone. The equal-time power
spectrum is recovered by setting $\chi=\chi'$.

Using the above definition, the angular two-point function of the Weyl
potential on two shells is
\begin{align}
\left\langle
\Phi_{{\rm W},\ell m}(\chi)
\Phi_{{\rm W},\ell' m'}^\ast(\chi')
\right\rangle
&=
(4\pi)^2 i^\ell(-i)^{\ell'}
\int\frac{\mathrm d^3 k}{(2\pi)^3}\,
j_\ell(k\chi)j_{\ell'}(k\chi')\,
Y_{\ell m}^{\ast}(\hat{\mathbf k})
Y_{\ell' m'}(\hat{\mathbf k})\,
P_{\Phi_{\rm W}}(k;\chi,\chi')
\nonumber\\
&=
\delta_{\ell\ell'}\delta_{mm'}\,
\frac{2}{\pi}
\int_0^\infty \mathrm d k\,k^2\,
j_\ell(k\chi)j_\ell(k\chi')\,
P_{\Phi_{\rm W}}(k;\chi,\chi') .
\end{align}
In the second line we used the orthonormality of the spherical harmonics in
$\hat{\mathbf k}$ and $(4\pi)^2/(2\pi)^3=2/\pi$.

We can now compute the convergence two-point function. From the expression
for $\kappa_{\ell m}$ above,
\begin{align}
\left\langle
\kappa_{\ell m}(z_s)
\kappa_{\ell' m'}^\ast(z_s)
\right\rangle
=
\ell^2(\ell+1)^2
\int_0^{\chi_s}\mathrm d\chi\,
W_{\phi^{\rm L}}(\chi)
\int_0^{\chi_s}\mathrm d\chi'\,
W_{\phi^{\rm L}}(\chi')\,
\left\langle
\Phi_{{\rm W},\ell m}(\chi)
\Phi_{{\rm W},\ell' m'}^\ast(\chi')
\right\rangle .
\end{align}
Substituting the shell two-point function gives
\begin{align}
\left\langle
\kappa_{\ell m}(z_s)
\kappa_{\ell' m'}^\ast(z_s)
\right\rangle
=
\delta_{\ell\ell'}\delta_{mm'}\,
\frac{2}{\pi}\,\ell^2(\ell+1)^2
&\int_0^\infty \mathrm d k\,k^2
\int_0^{\chi_s}
\frac{\mathrm d\chi}{\chi}\,
\frac{\chi_s-\chi}{\chi_s}
\int_0^{\chi_s}
\frac{\mathrm d\chi'}{\chi'}\,
\frac{\chi_s-\chi'}{\chi_s}
\nonumber\\
&\times
j_\ell(k\chi)
j_\ell(k\chi')
P_{\Phi_{\rm W}}(k;\chi,\chi') .
\end{align}
Comparing with the definition
\begin{equation}
\left\langle
\kappa_{\ell m}(z_s)
\kappa_{\ell' m'}^\ast(z_s)
\right\rangle
=
\delta_{\ell\ell'}\delta_{mm'}\,
C_\ell^{\kappa\kappa}(z_s),
\end{equation}
we recover Eq.~\eqref{eq:Cl_kappa_full_Phi_dollars2}.

The Limber approximation \citep{1953ApJ...117..134L, Bartelmann:1999yn}
gives a simpler form of Eq.~\eqref{eq:Cl_kappa_full_Phi_dollars2} when the
multipole is sufficiently large and the radial kernels and power spectrum
vary slowly compared with the oscillations of the spherical Bessel
functions. At large $\ell$, $j_\ell(k\chi)$ is strongly weighted around
\begin{equation}
k\chi\simeq \ell+\frac{1}{2}.
\end{equation}
Thus, for a fixed radial distance $\chi$, the projection mainly samples
Fourier modes near
\begin{equation}
k_\ell(\chi)
\simeq
\frac{\ell+1/2}{\chi}.
\end{equation}
The rapidly oscillating Bessel functions also suppress contributions from
widely separated radial distances. This is the sense in which the Limber
approximation makes the radial projection approximately local.

The exact localization identity underlying this approximation is the
spherical-Bessel orthogonality relation
\begin{equation}
\int_0^\infty \mathrm d k\, k^2\,
j_\ell(k\chi)\,j_\ell(k\chi')
=
\frac{\pi}{2\chi^2}\,
\delta_{\rm D}(\chi-\chi') .
\end{equation}
If the remaining factor multiplying the Bessel functions varies slowly over
the width of the Bessel kernel, it can be evaluated at the characteristic
scale $k_\ell(\chi)$ and at equal radial distances, $\chi'=\chi$. Therefore,
for a smooth function $P(k;\chi,\chi')$,
\begin{equation}
\int_0^\infty \mathrm d k\,k^2\,
j_\ell(k\chi)\,
j_\ell(k\chi')\,
P(k;\chi,\chi')
\simeq
\frac{\pi}{2\chi^2}
\delta_{\rm D}(\chi-\chi')\,
P\!\left(
\frac{\ell+1/2}{\chi};\chi,\chi
\right).
\end{equation}
This is the Limber replacement: the full unequal-time radial projection is
approximated by an equal-time contribution evaluated at the transverse
wavenumber $k\simeq(\ell+1/2)/\chi$.

Applying this replacement to
Eq.~\eqref{eq:Cl_kappa_full_Phi_dollars2} gives
\begin{equation}
C_{\ell}^{\kappa\kappa}(z_s)
\simeq
\ell^2(\ell+1)^2
\int_0^{\chi_s}
\mathrm d\chi\,
\frac{(\chi_s-\chi)^2}{\chi_s^2\,\chi^4}\,
P_{\Phi_{\rm W}}
\!\left(
\frac{\ell+1/2}{\chi};\chi,\chi
\right).
\label{eq:Cl_kappa_limber_PhiW_dollars2}
\end{equation}
This expression is useful for building intuition: high multipoles probe
smaller transverse scales, while the lensing kernel selects the relevant
range of distances along the line of sight. However, this intuition should
not be applied directly at low $\ell$, where the full-sky expression in
Eq.~\eqref{eq:Cl_kappa_full_Phi_dollars2} receives contributions from a
broader range of wavenumbers and radial distances.

The factor $\chi^{-4}$ in
Eq.~\eqref{eq:Cl_kappa_limber_PhiW_dollars2} should not be interpreted as a
divergence of the lensing efficiency near the observer. This form is written
in terms of the Weyl-potential power spectrum. Schematically, the Poisson
relation gives
\begin{equation}
\Phi_{\rm W}(k,\chi)
\propto
\frac{1}{k^2}\Delta_{\rm m}(k,\chi),
\end{equation}
so that
\begin{equation}
P_{\Phi_{\rm W}}(k;\chi,\chi)
\propto
\frac{1}{k^4}
P_{\Delta_{\rm m}}(k;\chi,\chi).
\end{equation}
In the Limber approximation,
\begin{equation}
\frac{1}{k^4}
=
\frac{\chi^4}{(\ell+1/2)^4},
\end{equation}
which cancels the apparent $\chi^{-4}$ factor in
Eq.~\eqref{eq:Cl_kappa_limber_PhiW_dollars2}. The endpoint divergence is
therefore an artefact of writing the spectrum in terms of
$P_{\Phi_{\rm W}}$ rather than in terms of the matter perturbation.

\bibliographystyle{JHEP}
\bibliography{Bibliography.bib}
\end{document}